\documentclass[10pt,journal,compsoc]{IEEEtran}
\usepackage{subcaption}
\usepackage{graphicx}
\usepackage[normalem]{ulem}
\usepackage{multirow}    
\usepackage{pifont}      
\usepackage{xcolor}       
\usepackage{xspace}      
\usepackage{flushend}    
\usepackage{upgreek}

\usepackage{times}
\usepackage{datetime}
\usepackage{url}
\usepackage{hyperref}
\hypersetup{
  colorlinks,
  linkcolor={red!50!black},
  citecolor={blue!50!black},
  urlcolor={blue!80!black}
}

\begin{document}

\newcommand{\name}{PiBooster\xspace}
\newcommand{\cache}{PiBooster cache\xspace}
\newcommand{\module}{PiBooster module\xspace}
\newcommand{\prename}{Pre-PiBooster\xspace}
\newcommand{\dynname}{Dyn-PiBooster\xspace}
\newcommand{\eat}[1]{}  
\newcommand{\authcomment}[3]{\textcolor{#3}{#1 says: #2}}\newcommand{\yueqiang}[1]{\authcomment{Yueqiang}{#1}{red}}
\newcommand{\zhi}[1]{\authcomment{Zhi}{#1}{blue}}
\newcommand{\mypara}[1]{\vspace{2pt}\noindent\textbf{{#1. }}}

\title{\Large \bf PiBooster: A Light-Weight Approach to Performance Improvements in Page Table Management for Paravirtual Virtual-Machines}
%
%
%

\author{Zhi~Zhang, Yueqiang~Cheng

\IEEEcompsocitemizethanks{
\IEEEcompsocthanksitem 
Zhi Zhang is with the Data61, CSIRO, Australia and the University of New South Wales, Australia.\protect\\
E-mail: zhi.zhang@data61.csiro.au 

\IEEEcompsocthanksitem Yueqiang Cheng is with Baidu XLab, America.\protect\\
E-mail: chengyueqiang@baidu.com 
}}
\IEEEtitleabstractindextext{%
\begin{abstract}
In paravirtualization, the page table management components of the guest operating systems are properly patched for the security guarantees of the hypervisor.
However, none of them pay enough attentions to the performance improvements, which results in two noticeable performance issues.
First, such security patches exacerbate the problem that the execution paths of the guest page table (de)allocations become extremely long, which would consequently increase the latencies of process creations and exits.
Second, the patches introduce many additional IOTLB flushes, leading to extra IOTLB misses, and the misses would have negative impacts on I/O performance of all peripheral devices.

In this paper, we propose \name, a novel lightweight approach for improving the performance in page table management.
First, \name shortens the execution paths of the page table (de)allocations by the \cache, which maintains dedicated buffers for serving page table (de)allocations.
Second, \name eliminates the additional IOTLB misses with a fine-grained validation scheme, which performs page table and DMA validations separately, instead of doing both together.
We implement a prototype on Xen with Linux as the guest kernel. We do small modifications on Xen ($166$ SLoC) and Linux kernel ($350$ SLoC).
We evaluate the I/O performance in both micro and macro ways.
The micro experiment results indicate that \name is able to \emph{completely eliminate} the additional IOTLB flushes in the workload-stable environments, and effectively reduces (de)allocation time of the page table by 47\% on average.
The macro benchmarks show that the latencies of the process creations and exits are expectedly reduced by 16\% on average.
Moreover, the \emph{SPECINT}, \emph{lmbench} and \emph{netperf} results indicate that \name has \emph{no} negative performance impacts on CPU computation, network I/O, and disk I/O.
\end{abstract}

\begin{IEEEkeywords}
Pare-virtualization, Peripheral Devices, Page-table Management.
\end{IEEEkeywords}
}

\maketitle


%
\IEEEpeerreviewmaketitle

\section{Introduction} \label{sec:intro}
In paravirtualization~\cite{XEN-SOSP03,whitaker2002scale}, the operating system of each Virtual Machine (a.k.a. guest or guest domain) and the hypervisor share the same virtual space.
In order to prevent malicious accesses from the guest OS, the hypervisor sets the guest page tables read-only, and validates their updates to ensure that there is no runtime violation~\cite{XEN-SOSP03}.
However, only the page table based protection is not enough to defend against the DMA attacks driven by the malicious guest OS~\cite{disaggregation}.
To fix this gap, the hypervisor resorts to the I/O virtualization (AMD-Vi~\cite{amdvt} or Intel VT-d~\cite{intelvt}) technology, which leverages a new Input/Output Memory Management Unit (IOMMU) to restrict DMA accesses to the physical memory pages occupied by the hypervisor and the guest page tables.
To integrate the above protection techniques, the guest page table management and the hypervisor are required to be properly patched.

\begin{figure*}[ht]
\centering
\includegraphics[width=0.9\textwidth]{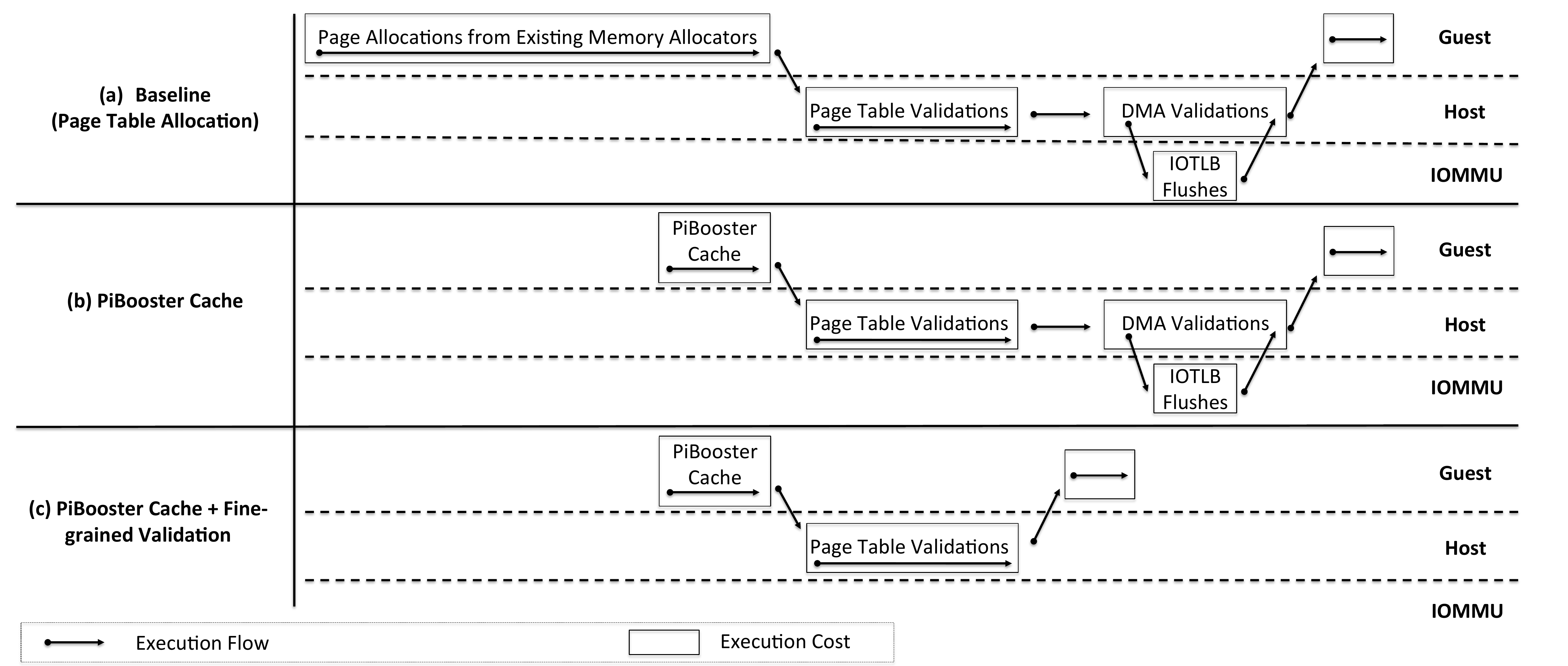} \\
\caption{Solution Overview. When the \name cache and fine-gained validation scheme is enabled,
the execution path of page table allocation is dramatically reduced, and the additional IOTLB flushes are eliminated.}
\label{fig:overview}
\end{figure*}

\mypara{Problem}
However, all existing patches mainly focus on the security enhancements of the hypervisor, without enough attentions to the performance improvements, which results in two noticeable performance issues.
The first one is the long execution paths of the guest page table (de)allocations, which involve a complex memory allocation process and an additional security validation procedure.
The memory allocation process frequently involves a \emph{slab} allocator~\cite{slaballocator} and the page frame allocations that are frequently managed with a buddy system~\cite{buddyallocator}, which introduces deep invocations for each page-table (de)allocation.
Moreover, the additional security validation procedure always adds extra costs for preventing page tables from malicious software and DMA modifications.
All these lead to poor performance of the page table (de)allocation, and consequently result in the long latencies of the creations and exits of processes.

The other one is the additional IOTLB flushes introduced by the security validations of the page table (de)allocations.
The guest page tables should be non-readable for any DMA requests and the corresponding pages should be readable and writable when the page tables are deallocated.
The updates of the access permissions require IOTLB flushes to refresh the access permissions, which is necessary for the sake of the security of the hypervisor.
In addition, these access permission update events are \emph{often} triggered during the whole life cycle of a running system.
As a consequence, the IOTLB flushing events are \emph{frequently} involved, which inevitably increases the IOTLB miss rate and lowers the speed of the DMA address translation.
All these are likely to introduce negative impacts on the I/O performance of all peripheral devices.
The baseline of Figure~\ref{fig:overview} (a) illustrates the two issues taking page table allocation as an example.

\mypara{Solution}
The above identified performance issues urge us to revise the design of the page table management to improve the performance without sacrificing the security guarantees.
In response to the appeal, in this paper we propose \name, a novel software-only lightweight approach for improving performance in page table management.
First, \name shortens the execution paths of the page table (de)allocations by the \cache, which maintains dedicated buffers for serving page table (de)allocations (Figure~\ref{fig:overview} (b)).
The \cache queues the deallocated page-table pages in a constant time, with the hope that
they will be reused (popped out of the cache) in the page table allocations in the near future.
By doing so, the page table allocations do not need to involve the costly memory management subsystem every time, instead they could directly get pages from the cached buffers, dramatically shortening the execution paths.

Second, \name eliminates the additional IOTLB flushes with a fine-grained validation scheme (Figure~\ref{fig:overview} (c) ), which separates the page table and DMA validations.
In the traditional design, there are two types of pages: writable page that is writable for software and DMA accesses, and non-writable page (e.g., page-table page) that are non-writable for both software and DMA.
The page table allocations and deallocations always involve the type changes between the two types.
Thus, the hypervisor has to do both page table and DMA validations to ensure that both of them are not violating the security policies.
However, we observe that it is not necessary to do DMA validation every time if we create a new page type (i.e., semi-writable page) with the non-writable permission for DMA access, and enforce the page type changes only occur between the page-table pages and the semi-writable pages during the page table allocations and deallocations.
The management of the semi-writable pages can be assisted by the \cache.

We implement a prototype on Xen with Linux as the guest kernel. We do small modifications on Xen version 4.2.1 ($166$ SLoC) and Linux kernel version 3.2.0 ($350$ SLoC).
We evaluate the I/O performance in both micro and macro ways.
The micro experiment results indicate that \name is able to completely eliminate the additional IOTLB flushes, and effectively reduce (de)allocation time of the page table.
There are (34\%, 47\%), (38\%, 22\%) and (65\%, 65\%) improvements for a pair of allocation and deallocation for three-level page table, from top to bottom.
Even in the worst cases when the \name has to go through the traditional path to allocate pages, the performance overhead for one page table allocation is still very small, only adding about 20 instructions. 
Fortunately, the worst case could be avoided by carefully setting the number of the initial pages of the \cache.
The macro benchmarks show that \name has no negative impact on the CPU computation, network I/O and disk I/O.
In particular, the latencies of the process creations and exits are expectedly reduced by 16\% on average.

In summary, we make the following contributions:
\begin{enumerate}
\item We identify two significant performance issues in the page table management. In particular, we are the first, to the best of our knowledge, to identify the performance issue between guest page table (de)allocations and the IOTLB flushes.
\item We proposed a novel approach - called \name, to shorten the execution paths of the page table allocation and deallocations, and eliminate the additional IOTLB flushes, without sacrificing the system security.
\item We implemented a prototype of the \name and evaluated the performance in both micro and macro ways. The experiment results indicate that the \name can benefit the page table (de)allocations and eliminate additional IOTLB flushes, without negative performance impacts on the system.
\end{enumerate}

The rest of the paper is structured as follows: In Section~\ref{sec:prob}, we briefly describe the background knowledge, and highlight the performance issues. Then we describe the system overview and implementation in Section~\ref{sec:overview} and Section~\ref{sec:impl}. In Section~\ref{sec:eva}, we evaluate the performance of the \name system, and discuss several issues in Section~\ref{sec:dis}. At last, we discuss the related work in Section~\ref{sec:related}, and conclude the whole paper in Section~\ref{sec:con}.

\section{Problem Definition} \label{sec:prob}
In this section, we describe the necessary background knowledge and highlight the identified two performance issues: 1) long execution paths of the guest page table (de)allocations and 2) the additional IOTLB flushes.
As Xen~\cite{XEN-SOSP03} is a typical and popular paravirtual hypervisor, we use Xen in the x86 MMU model~\cite{x86-pv-model} to illustrate the details.
The presented mechanisms are also available on other paravirtual platforms.

\subsection{Long Execution Path}\label{sec:longpath}
The long execution path issue refers to the execution paths of the guest page table allocations and deallocations.
In the traditional design, allocating and deallocating a page-table page are time costly, as they have to invoke the complex memory allocators and perform both software (i.e., page table) and DMA validations.
In addition, such allocation and deallocation events frequently occur in a system, i.e., the numerous process creations and exits will result in many page-table allocations and deallocations.
Thus, it is necessary to deeply understand the long execution path issue and improve the performance accordingly.

\subsubsection{Page Allocation and Deallocation}
To allocate a page, the guest kernel has to invoke the system allocators, typically from the slab allocator to the buddy allocator.
The slab allocator~\cite{slaballocator} manages a variable number of caches that are linked together by a doubly linked circular list.
Each cache maintains blocks of contiguous pages in memory called slabs, which are carved up into small chunks for data structures and objects of specific sizes.
When \emph{kmalloc} is called, the allocator will search through the prepared caches.
If there is no suitable object, the buddy allocator will be involved.
The buddy allocator~\cite{buddyallocator} manages all free memory blocks of permitted sizes by free lists. The blocks consist of pages and the sizes are usually powers of 2.
When receiving requests for memory, the allocator firstly checks if the request size is equal to a permitted size. If they are equal, the allocator will return a block from the corresponding free list. If they are not equal or the free list is empty, the allocator will split a larger block (e.g., the permitted size is twice that of the requested size) into multiple sub-blocks and return one sub-block, inserting the rest to the proper free list.

In contrast to the page allocation, the page deallocation is to return the page back to the system.
The deallocation process may also invoke slab allocator and/or buddy allocator, and would trigger the updates of the corresponding data structures,
e.g., the buddy allocator always attempts to merge adjacent freed blocks into a large permitted size.

In brief, the page allocation and deallocation are time costly due to the deep invocations and complex updates of the dependent data structures.

\subsubsection{Page Table Validations}\label{sec:pv-security}
Page tables are used by the hardware, i.e., Memory Management Unit (MMU), to translate the linear addresses into physical addresses.
In the PAE-enabled paging mode, a page table has three levels: L3 level (bottom level), L2 level (middle level) and L1 level (top level).
The slots in L3, L2 and L1 levels are known as Page Table Entry (PTE), Page Middle Directory (PMD) and Page Global Directory (PGD), respectively.
A PTE slot could determine the access permissions of a page, e.g., the kernel could set a page read-only by clearing the bit within a PTE slot that represents the writable permission.
Typically, each user process has its own page table, and the creation and exit of a user process will be accompanied by the allocation and deallocation of a page table respectively.
\begin{figure}[ht]
\centering
\includegraphics[width=0.45\textwidth]{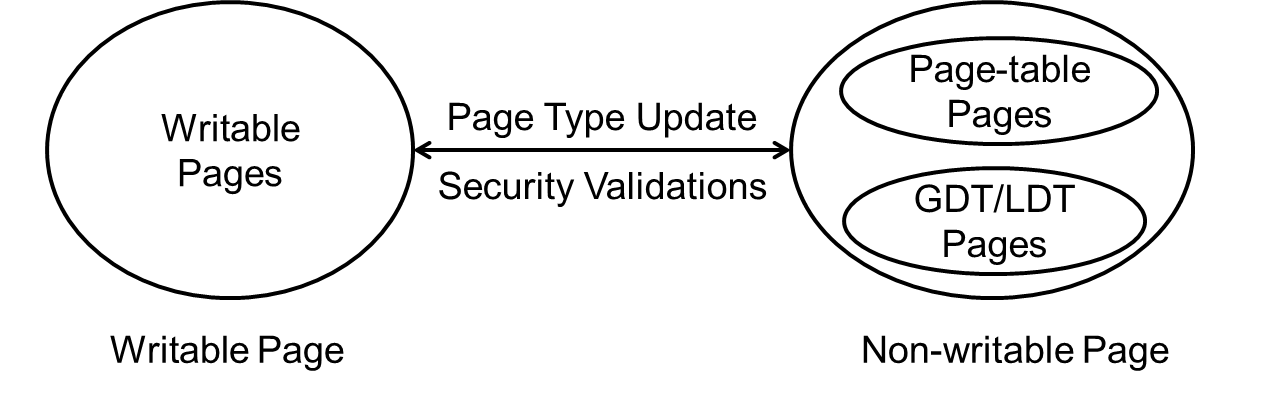} \\
\caption{Page type updates between writable pages and non-writable pages. }
\label{fig:page-type-updates}
\end{figure}


The hypervisor defines two page types 1) \emph{writable page} that is writable for both software and DMA, and 2) \emph{non-writable page} that is non-writable for both software and DMA.
The non-writable page has several sub-page types, such as page-table pages and GDT/LDT pages.
In addition, the hypervisor also requires that every page type update only occur between writable and non-writable pages, as summarized in Figure~\ref{fig:page-type-updates}.
The hypervisor maintains a type reference count for each page, and enforces the policy that any given page has exactly one type at any given time.
In addition, it also enforces that only pages with the writable type have a writable mapping in the page tables.
By doing this it can ensure that the guest OS is not able to directly modify any page-table pages and therefore cannot subvert the security of the hypervisor.

Whenever a page table is loaded to work, there should be some page type updates from writable pages to non-writable pages (i.e., page-table page).
First, the hypervisor must ensure that the page-table page has a type reference count of zero.
In addition, it must be validated to ensure that it complies with the following policy:
for a page with a page-table type to be valid, it is required that any pages referenced
by a present page table entry in the page have the type of the next level down.
For instance, any page referenced by a page with type L1 Page Table must itself have the type L2 Page Table.
This policy is applied recursively down to the L3 page table layer.
At L3 the invariant is that any data page mapped as writable by an L3 entry must be a writable page.
By applying these policies, the hypervisor ensures that all page-table pages as a whole are safe to be loaded.
Note that the hypervisor is always involved in all updates of the page tables, the policies for the page table updates are non-bypassable.

The validation process for the page table deallocation is contrary to the one for the page table allocation.
In brief, the hypervisor will validate the page table's type reference count, access permissions and the slot mappings to ensure that the page table is completely and securely deallocated.

\subsection{DMA Validations and IOTLB Flushes}
\subsubsection{DMA Address Translation}
The input/output memory management unit (IOMMU)~\cite{intelvt} is a memory management unit (MMU) that connects a DMA-capable I/O bus to the main memory.
Like a traditional MMU, the IOMMU maps device addresses (also called as I/O addresses) to physical addresses through a dedicated page table.
The IOMMU page table that is created and maintained by the hypervisor in its own space is able to restrict the access to a particular page by configuring the permission bits.
The hypervisor grants different access permissions to different page types, such as the writable pages are always allowed with full access permissions, while the page-table pages are always inaccessible to any devices.

\begin{figure}[!t]
    \begin{subfigure}{0.45\textwidth}
        \includegraphics[width=1\textwidth]{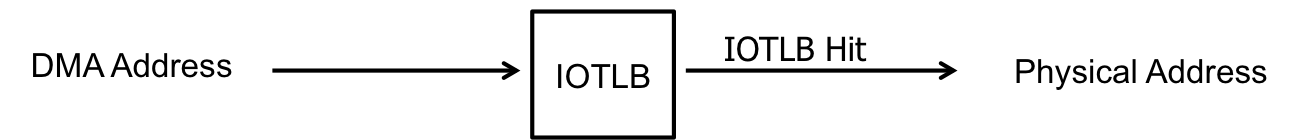}
        \caption{\centering IOTLB Path.}
        \label{fig:iotlbpath}
    \end{subfigure}
    \vfill
    \begin{subfigure}{0.45\textwidth}
        \includegraphics[width=1\textwidth]{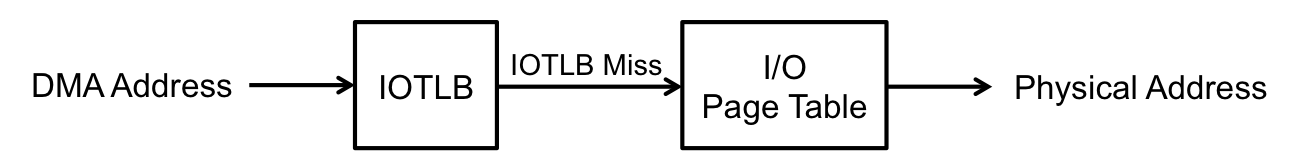}
        \caption{\centering I/O Page-Table Path.}
        \label{fig:ioptpath}
    \end{subfigure}
    \caption{IOTLB path is much faster than I/O page-table path.}
    \label{fig:dma-add-trans}
\end{figure}

However, if the DMA address translation always needs to look up the IOMMU page table, it will be slow and inefficient.
To accelerate the translation speed, the I/O translation look-aside buffer (IOTLB) is introduced.
The IOTLB is used to cache frequently accessed page table entries.
By doing so, the IOTLB is very likely to be accessed, indicating that the physical address of a queried DMA address will be immediately fetched through the IOTLB path (Figure~\ref{fig:iotlbpath}).
If unlikely the IOTLB miss occurs, the DMA address translation still can go along the slow I/O page-table path to get the physical address (Figure~\ref{fig:ioptpath}).
To achieve a better I/O performance, the DMA address translation should avoid taking the I/O page-table path.

\subsubsection{Additional IOTLB Flushes}
The page table allocation and deallocation always trigger the updates between the writable pages and the page-table pages, and they have different access permissions for DMA requests.
To keep the security of the hypervisor, the DMA validation is necessary.
Specifically, the hypervisor will update the corresponding entries of the IOMMU page table to set correct access permissions.
After that, the hypervisor also needs to flush IOTLB entries to invalidate the stale entries, without which the security of the hypervisor would be violated, e.g., the DMA requests could write the page-table pages through stale IOTLB entries.

\begin{figure}[ht]
\centering
\includegraphics[width=0.45\textwidth]{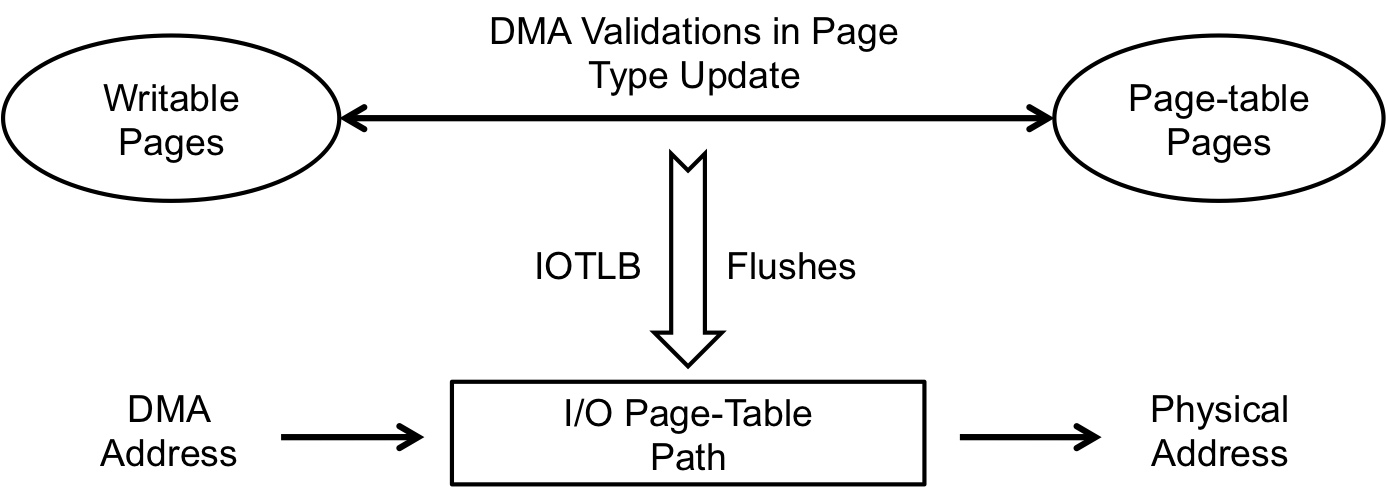} \\
\caption{The DMA validations introduce additional IOTLB flushes, which lead to selecting the slow path  (i.e., I/O page-table path) for the DMA address translation.}
\label{fig:pro-ill}
\end{figure}

In addition, the additional IOTLB flushes are likely to let the DMA address
translations take the slow and inefficient page-table path,
instead of taking the fast and efficient IOTLB path (Figure~\ref{fig:pro-ill}), due to the additional IOTLB misses.
Numerous IOTLB misses would lower the speed of the DMA transferring, especially for the high performance devices, such as Intel I/OAT ~\cite{lauritzenintel}.


\section{\name Overview} \label{sec:overview}
\begin{figure*}[ht]
\centering
\includegraphics[width=0.8\textwidth]{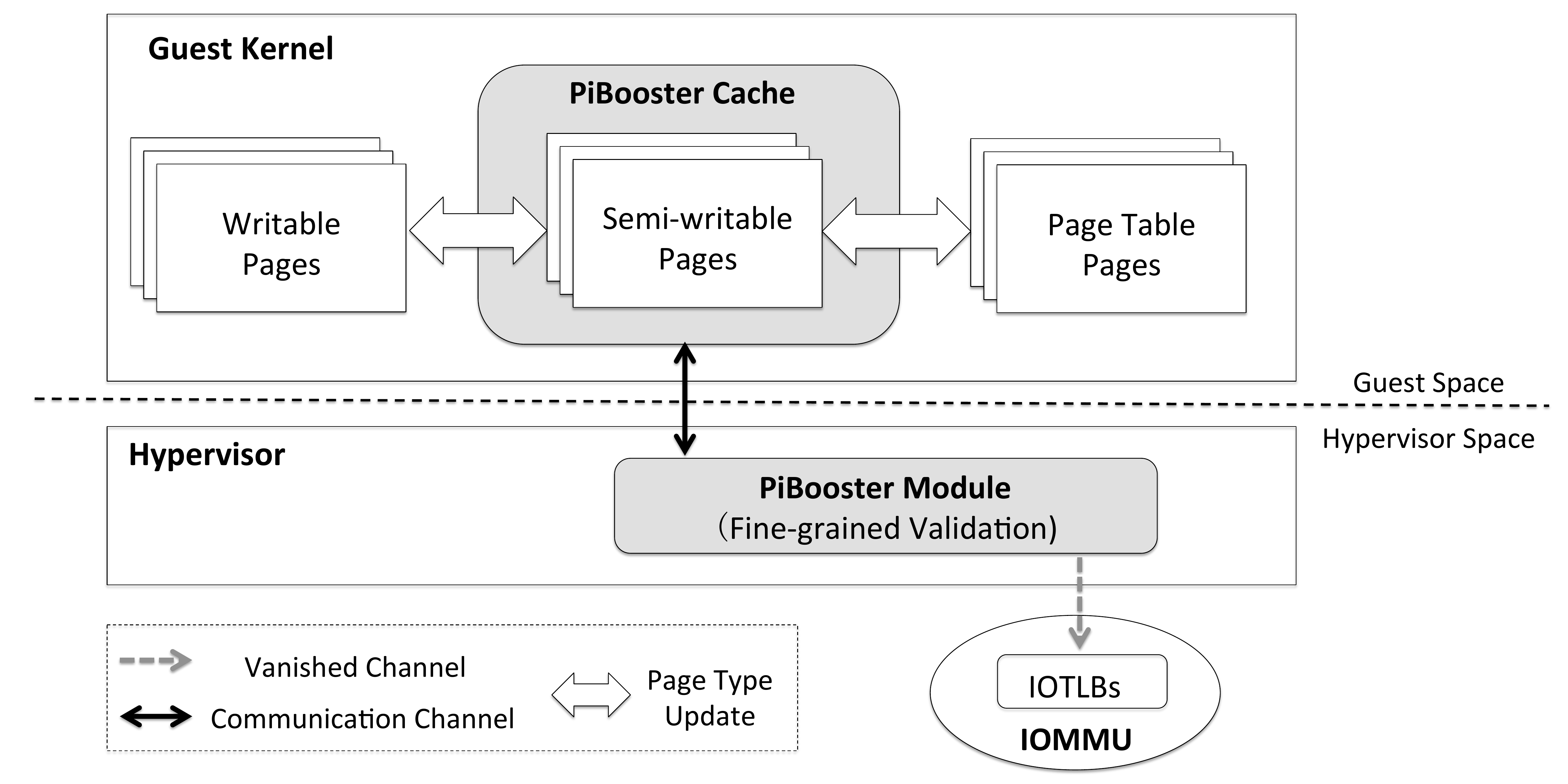} \\
\caption{\name Architecture. The semi-writable pages are managed by the \name cache, and all page type updates are validated by the \name module.
Through the cooperation of \name cache and \name module, \name successfully shortens the execution paths of the guest page table (de)allocations and eliminates additional IOTLB flushes.}
\label{fig:arch}
\end{figure*}

\subsection{Design Requirements}\label{sec:req}
In the design of \name, we consider several requirements, which are summarized and listed as follows.
\begin{enumerate}
\item Unaltered system security. The new scheme should not sacrifice the system security to obtain performance benefits. No one likes to use a system with known design loopholes.
\item Compatible with legacy applications. The new scheme should limit the modifications on the guest kernel and the hypervisor, without any modifications on the existing applications.
\item Small modifications. The new scheme should minimize the development cost on the guest kernel and the hypervisor.
\end{enumerate}

\subsection{\name Architecture}
Figure~\ref{fig:arch} depicts the architecture of \name. It consists of a \cache in the guest kernel and a \name module in the hypervisor space.
The \cache allocates a number of semi-writable pages at the initial stage,  which are newly introduced by the fine-grained validation scheme (Section~\ref{sec:fine-grained}).
These semi-writable pages are maintained in a dedicated cache.
At runtime, the \cache attempts to satisfy all page table allocation and deallocation requests, shortening the execution path and saving the execution time.
There are many page type updates from/to the semi-writable pages.
The \cache mediates these updates and issues hypercalls to the hypervisor, asking it to perform the fine-gained security validations on the update requests (i.e., the communication channel).
Once the verified pages pass the validations, their page types will be updated accordingly.

The \module enforces the fine-grained validation scheme on the page table updates.
It performs (1) the page table validations that validate the page table contents as well as the type reference count, and (2) the DMA validations that ensure that the DMA requests cannot write the page-table pages and the semi-writable pages.
In the traditional validation scheme, the DMA validations always trigger additional IOTLB flushes due to the access permission updates between the writable pages and the page-table pages.
In the fine-grained validation scheme, both semi-writable pages and page-table pages are non-writable for DMA requests.
Thus, the hypervisor never needs to do the IOTLB flushes or IOMMU page table updates (i.e., vanished channel), which will benefit the I/O performance of the peripheral devices.
Moreover, it also saves the time of the whole security validation, further reducing the execution time of the page table allocations and deallocations.

\subsection{Fine-Grained Validation}\label{sec:fine-grained}
The fine-grained validation scheme aims to eliminate the additional IOTLB flushes and reduces the total time of the security validations.
Specifically, in the traditional security validation scheme, there are only two general page types: writable page and non-writable page,
and the type updates between them are required to do both page table validations and DMA validations.
The DMA validations not only increase the total validation time, but also introduce numerous additional IOTLB flushes.
Moreover, the additional IOTLB flushes cannot be skipped, because it would provide a time gap for the adversary to attack the hypervisor by leveraging the stale IOTLB entries.

To address this problem without sacrificing the system security, we introduce a new page type: \emph{semi-writable page}, which is writable for software but non-writable for DMA.
In addition, we enforce that the page type updates between the writable page and the page-table page must go through the semi-writable page first (as illustrated in Figure~\ref{fig:arch}).
As the semi-writable page and page-table page are already inaccessible to DMA, there is no need to do the DMA validations, meaning that the additional IOTLB flushes could be totally avoided.
As a consequence, the time of the whole security validation process is reduced, accelerating the speeds of page table allocations and deallocations.
Similar to the management of the page-table pages, the hypervisor is only responsible for the final validation of the semi-writable page, leaving all other management operations for the \cache.
By doing this, we can keep the modifications as small as possible by reusing existing validation process and page-table management subsystem, and also retain the system security.

\subsection{\name Module}\label{sec:module}
The \module works in the hypervisor space, extended from the traditional coarse-gained validation module.
The first task of the \module is to add the support for the \emph{semi-writable page}.
Instead of adding a new data structure for marking the semi-writable pages, the \module chooses to reuse the existing one.
By doing so, the \module could reuse the existing interfaces.
In particular, we find that the page-type data structure could be extended by borrowing a bit from \emph{type reference count} for \emph{semi-writable page}.

The second task of the \module is to perform the fine-grained security validations for all page type update requests.
The main logic is to check page type first, and then determine to perform one validation or both of them.
Specifically, if the page type update occurs between writable page and semi-writable page, it will perform both page table and DMA validations.
However, if the page type update occurs between semi-writable page and page-table page, the \module only perform the page table validations, skipping the DMA validations.
Note that the page table validations are always necessary, as the semi-writable pages are writable for the guest OS, and the modifications from the untrusted guest OS could subvert the enforced security policies.


The \module also exports a new hypercall interface for the \cache to facilitate their communications.
Through the new interface, the \cache could explicitly invoke the \module to perform security validations on specific page type updates.

\subsection{\name Cache}\label{sec:cache}
The basic idea behind the \cache is to have caches of semi-writable pages available for page table allocations and deallocations.
Without the page-oriented \cache, the kernel will spend much of its time allocating, initializing and freeing page-table pages.
The slab allocator that is similar to the \cache is not used in our settings, due to the following reasons.
First, it is not aware of the page type or the security requirements, which would lead to unexpected crashes of the guest OS.
Adding such functionalities would introduce many internal checking operations that are necessary for other objects maintained by the slab allocator.
At least, those additional checking operations will lower the (de)allocation efficiency of other objects.
Second, the existing size-oriented management of the slab does not distinguish the page-table pages from other pages that have the same size, and the customization of this management mechanism would need lots of development costs, such as interface updates, internal data structure updates.
In addition, the related components that rely on the slab allocator may also be affected.
At last, adding the fine-grained validation mechanism only for one object (i.e., page-table page) will subvert the generality of the slab allocator.
Considering the above three reasons, we give up reusing the existing slab allocator and aim to build a dedicated one, maintained by \cache, serving the page table allocations and deallocations.

\subsubsection{\name Cache Initialization and Destruction}
The \cache is enabled in the system bootup phase by default.
By doing so, the page tables of all user processes are served by the \cache from the very beginning.
To increase the flexibility, it also allows dynamical activation at runtime through an exported interface.

In the initialization phase, the \cache allocates a bulk of pages from the existing system allocators, converts them into semi-writable pages, and maintains them in a dedicated cache list.
At runtime, the page table is always successfully deallocated by efficiently pushing deallocated pages into the \cache.
But the cases for the page table allocation is a little bit complex.
Normally, the \cache could serves all page table allocation requests using the cached pages.
However, in the worst cases, the cached pages may not be able to satisfy the allocation requests.
In such conditions, the \cache will have to re-invoke the system allocators to get new pages.
Fortunately, the reinvocations could be avoided by carefully setting the number of the initial pages.
In fact, the re-invocations rarely occur in a workflow stable system.
In our experiments, there are always several  semi-writable pages in the \cache ready for the page-table allocations after the \name is running for a few minutes.
All these cases are evaluated in Section~\ref{sec:eva}.

The \cache works in the whole life cycle of the guest by default, but the end user is able to explicitly disable it at any time through the exported interface.
Once the \cache receives the \emph{disable} command, it will release all resources, e.g., deallocating the cached pages assisted by the existing system allocators as well as the \module, and releasing the data structures that are used for managing the cached pages.
In addition, it will also issue a hypercall to inform the hypervisor, which will disable the fine-grained validation mechanism and switch it back to the traditional coarse-grained validation scheme.

\subsubsection{Cache Shrinking}
When the memory management daemon (e.g., \emph{kswapd}) notices that the available memory is tight, it will explicitly call the exported interfaces of the \cache to free some cached pages.
There are two interfaces to shrink the cache pages. One is based on the page number. The memory management daemon can specify a number to ask the \cache to release.
The other one is based on the percentage. For instance, the kernel could ask the \cache to release 50\% cached pages.
In fact, the number of semi-writable pages maintained in the \cache is not too high. In our experiments, it is always less than $180$, meaning that the size of the cache is less than $720KB$.

The \cache could also automatically shrink itself through a predefined threshold.
The threshold can be defined according to the page number or the proportion (i.e., the number of the cached semi-writable pages over the number of the page-table pages), or a combination of them.

\section{Implementation} \label{sec:impl}
In this section, we present the implementation details of the \module and the \cache based on Xen (the hypervisor) and Linux (the guest kernel).

\subsection{\name Module}
\begin{figure}[ht]
\centering
\includegraphics[width=0.45\textwidth]{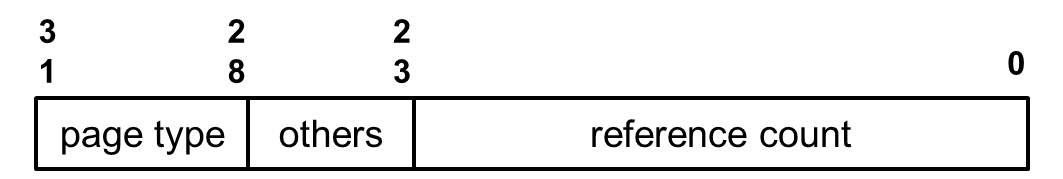} \\
\caption{The layout of the traditional data structure for page type.}
\label{fig:field-of-page-type-info}
\end{figure}
\begin{figure}[ht]
\centering
\includegraphics[width=0.45\textwidth]{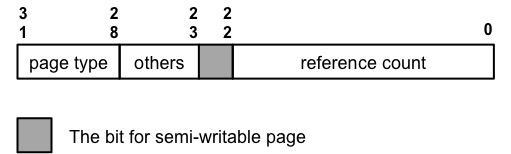} \\
\caption{The semi-writable page type support.}
\label{fig:field-of-semi-type}
\end{figure}

The first main task of the \module is to extend the existing data structure to support semi-writable page.
As illustrated in Figure~\ref{fig:field-of-page-type-info}, the data structure for labelling page types occupies 32bits, i.e., bits 28 - 31 are allocated for page type, bits 23 - 27 is for others (e.g., bit 26 indicate if this page has been validated), and bits 0 - 22 are for reference count of one page type.
The existing page types have occupied all page type bits, and there is no extra bit available for semi-writable page.
Facing this problem, we do not choose to introduce new data structures, as it would increase the management complexity and result in many modifications of all related management functions.
Instead, we choose to borrow a bit from \emph{reference count}.
In particular, the \emph{reference count} field has 23 bits, recording the number of type references for a page as its current type.
In fact, the system usually does not build so many references (i.e., $2^{23}-1$) to one page. Thus, we borrow the highest bit (bit 22) as the semi-page table bit (as illustrated in Figure~\ref{fig:field-of-semi-type}).
As a consequence, it still supports more than 4 million reference counts, enough for almost all cases. And the hypervisor will always do a check for the \emph{reference count} whenever doing operations on it so as to avoid count overflows. 
Actually, the hypervisor with \name is functioning well in our experiments.

The \module also needs to patch the page type checking functions, e.g., \_\_get\_page\_type, to adjust the checking logic.
The added checking logic is straightforward, such that we only add or change 166 SloC to achieve the whole patch.
In addition, some validation steps (e.g., DMA validations) necessary in the traditional design could be skipped when the newly introduced semi-writable page is involved in the page type update, simplifying the whole validation logic in a certain level.

\subsection{\name Cache}
There are three-level of the guest page table, and the \cache maintains a single-linked list for each of them.
Each node of the list has a pointer pointing to the next node and a page ID that is the base address of a cached page.
Note that this address should be a physical address, rather than a virtual address, because physical address of a page is unique in the whole system but its virtual addresses could be multiple.
Thus, using virtual address would lead to the confusion of the page type tracing and the semi-writable page management.
As the page table allocations and deallocations could happen at any time on any core, each list has its own lock to support concurrent updates.

The \cache has two interfaces for the runtime page table allocations and deallocations.
The \emph{pop} interface is for the page table allocations.
When this interface is invoked, the \cache fetches the top node of the corresponding list, extracts the base address of the cached page, and then returns it to the caller.
Correspondingly, the \emph{push} interface is serving the page table deallocations.
In the \emph{push} interface, the \cache saves the base address of the deallocated page into a node, and inserts it onto the top of the list.
Obviously, the \emph{pop} and \emph{push} interfaces are extremely fast as they could response the requests in a constant time.
In addition, the \cache also exports an interface for the memory management daemon for explicitly shrinking the cached pages.

The \cache also adds several virtual files in \emph{sysfs}~\cite{love2004linux,mochel2005sysfs} that is a virtual file system provided by the Linux kernel.
By using the virtual files, the end user could send commands to the \cache,  as well as query and configure the internal status.
In the current implementation, we only support one command, which is able to activate and deactivate the cache service in an on-demand way.
In addition, the end user could read and write the virtual files to query the number of the cached pages, and dynamically shrink the cache size.
For instance, the end user could explicitly ask the \cache to release all the cached pages by setting the number of the cached pages to be zero.

The \cache mediates all page type updates related to the semi-writable pages.
For each page type update, the \cache would issue the hypercall exported by the \module to explicitly inform the hypervisor to perform security validations.
In certain cases, such as shrinking the cached semi-writable pages to writable pages, the \cache could submit a batch of requests in one hypercall.
In such case, the hypervisor would update the I/O page tables as well as flush IOTLBs at one time, saving time of the cache shrinking.


\section{Evaluation} \label{sec:eva}
We have implemented the prototype of \name on our experiment platform.
Xen version 4.2.1 is the hypervisor while the guest VM (i.e., Dom0) is Ubuntu version 12.04 with Linux kernel version 3.2.0.
The \name added or changed $350$ SLoC in the Linux kernel and $166$ SLoC in Xen.
To fully evaluate the performance and its effects on the whole system, we measured the \name in both micro-benchamrks (e.g., the frequency of IOTLB flushes, the execution time of page table (de)allocations, and the memory usage of the \cache) and macro-benchmarks (e.g., \emph{SPECINT}, \emph{netperf} and \emph{lmbench}).

\subsection{Experiment Setting}
The experiment platform is a LENOVO QiTian M4390 PC with four CPU cores (i.e., Intel Core i5-3470) running at 3.20 GHz.
We enable the Intel VT-d feature through BIOS and grub configuration file. 


\mypara{Workload Emulation}
In order to allow us to repeatedly measure the effects of the \name on 1) page table allocations and deallocations, and 2) the IOTLB flushes, we use a stress tool to explicitly emulate a heavy workload with many short-time and concurrently-running processes.
Specifically, the tool periodically launches a browser (i.e., Mozilla Firefox 31.0 in the experiment), continuously opens new tabs one by one, and terminates the browser gracefully.
The purpose of these operations is to frequently create and terminate a large number of processes, leading to many page table allocations and deallocations.
The frequency can be configured. In our experiment setting, there are $542$ processes created and exited per minute.
In order to avoid the browser occupying too much memory, we terminate it in every 5 minutes.
At this moment, the memory usage of the browser reaches to $284.1$ MB on average.


\subsection{Micro-Benchmarks}
The micro-benchmark measurements are to evaluate the frequency of the IOTLB flushes, the execution time of the page table (de)allocations and the memory usage of the \cache.
For each measurement, there are two control groups and one baseline/normal group.
In the baseline group, we run the workload emulation in the guest VM with default settings, without enabling the \name mechanism.
On the contrary, the two control groups are: 1) the \prename group, where the \name is enabled before the workload emulation starts, and 2) the \dynname group, where the \name is dynamically enabled (e.g., five minutes after the workload emulation launches). The \dynname group is to evaluate if 1) the \name is able to enter a stable state as the \prename group, and 2) how fast the \name is able to enter the stable state.

\begin{figure}[ht]
\centering
\includegraphics[width=0.5\textwidth]{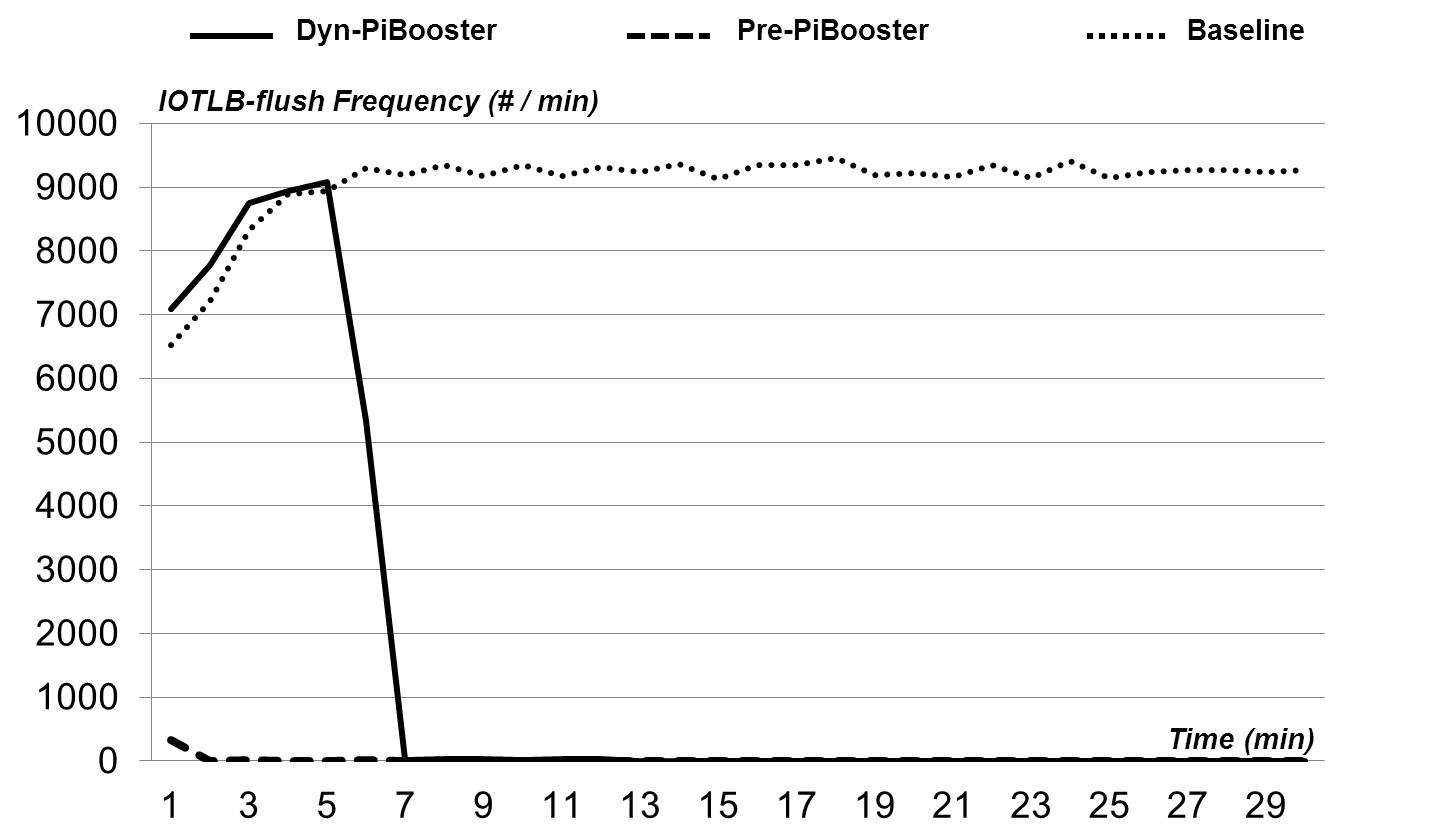} \\
\caption{The frequency of IOTLB flushes.  In the \prename group, the frequency is reduced from a very low level to zero within 1 minutes. In the \dynname group, the frequency drops sharply within two minutes from the high level to zero. Both control groups indicate that the \name could always enter the stable state (i.e., zero frequency).}
\label{fig:iotlbflush}
\end{figure}

\subsubsection{Frequency of IOTLB Flushes}
In this test, we aim to evaluate the effectiveness of the fine-grained validation on the additional IOTLB flushes.
We sample the frequency of the IOTLB flushes in 30 minutes.
The measurement results are illustrated in Figure~\ref{fig:iotlbflush}.
In the baseline group, the frequency of the IOTLB flushes is increasing in the first five minutes, and then keeps at a high flush rate (i.e., 9050 flushes on average per minute) until the test completes.
In the \prename group, the flush frequency quickly decreases from a low level (i.e., 332 flushes for the first minute) to zero level in about one minute and keeps at the level.
In the \dynname group, the flush frequency sharply decreases to zero when the \name is enabled. The \name roughly spends two minutes entering the stable state.
Thus, we can conclude that the fine-grained validation scheme is able to efficiently and effectively eliminate the IOTLB flushes introduced by the DMA validations.


\begin{figure*}[t!]
    \centering
    \begin{subfigure}[t]{0.5\textwidth}
        \centering
        \includegraphics[height=2.0in]{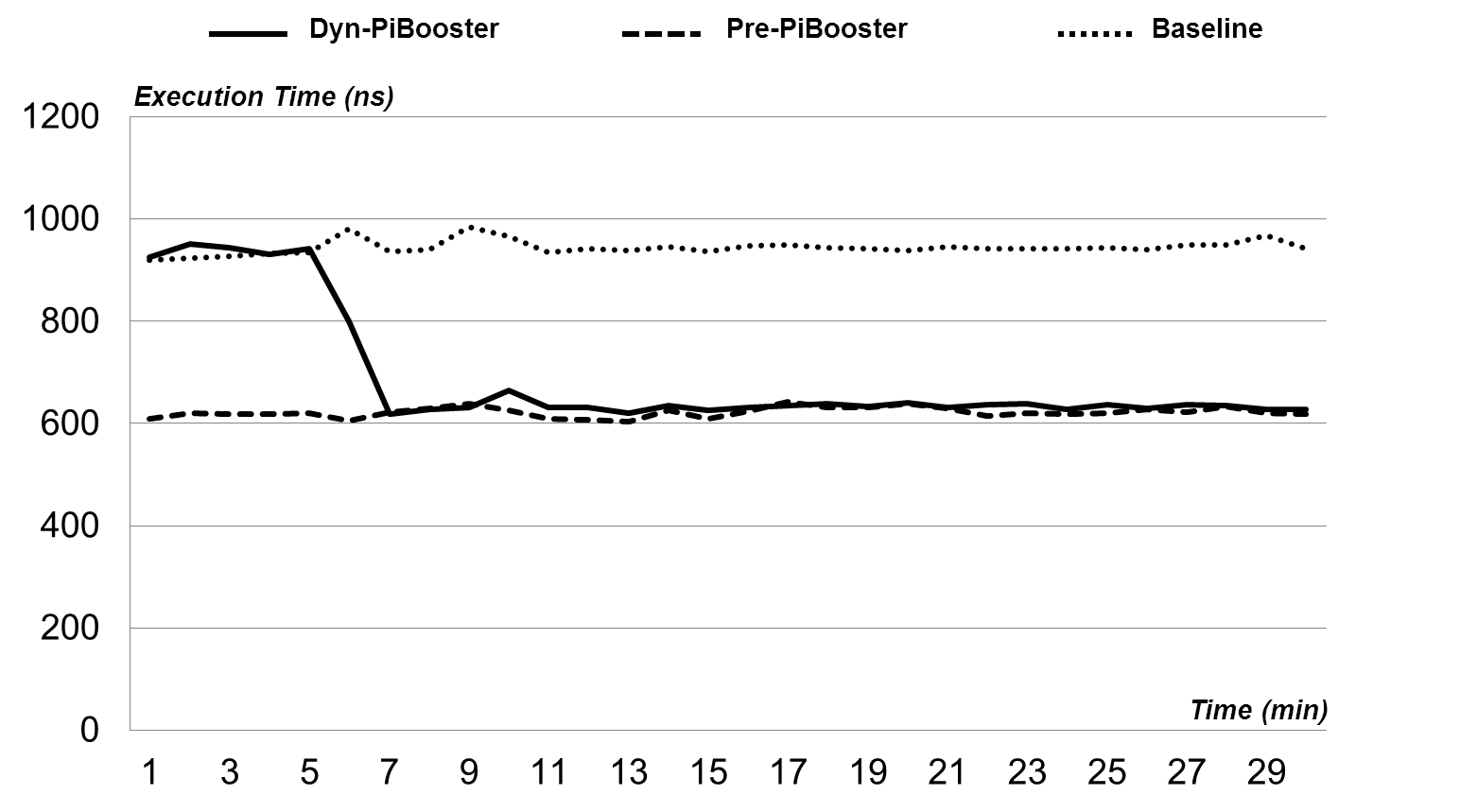}
        \caption{Execution time of L1 allocation is reduced by 34\%.}
        \label{fig:l1a}
    \end{subfigure}%
    ~
    \begin{subfigure}[t]{0.5\textwidth}
        \centering
        \includegraphics[height=2.0in]{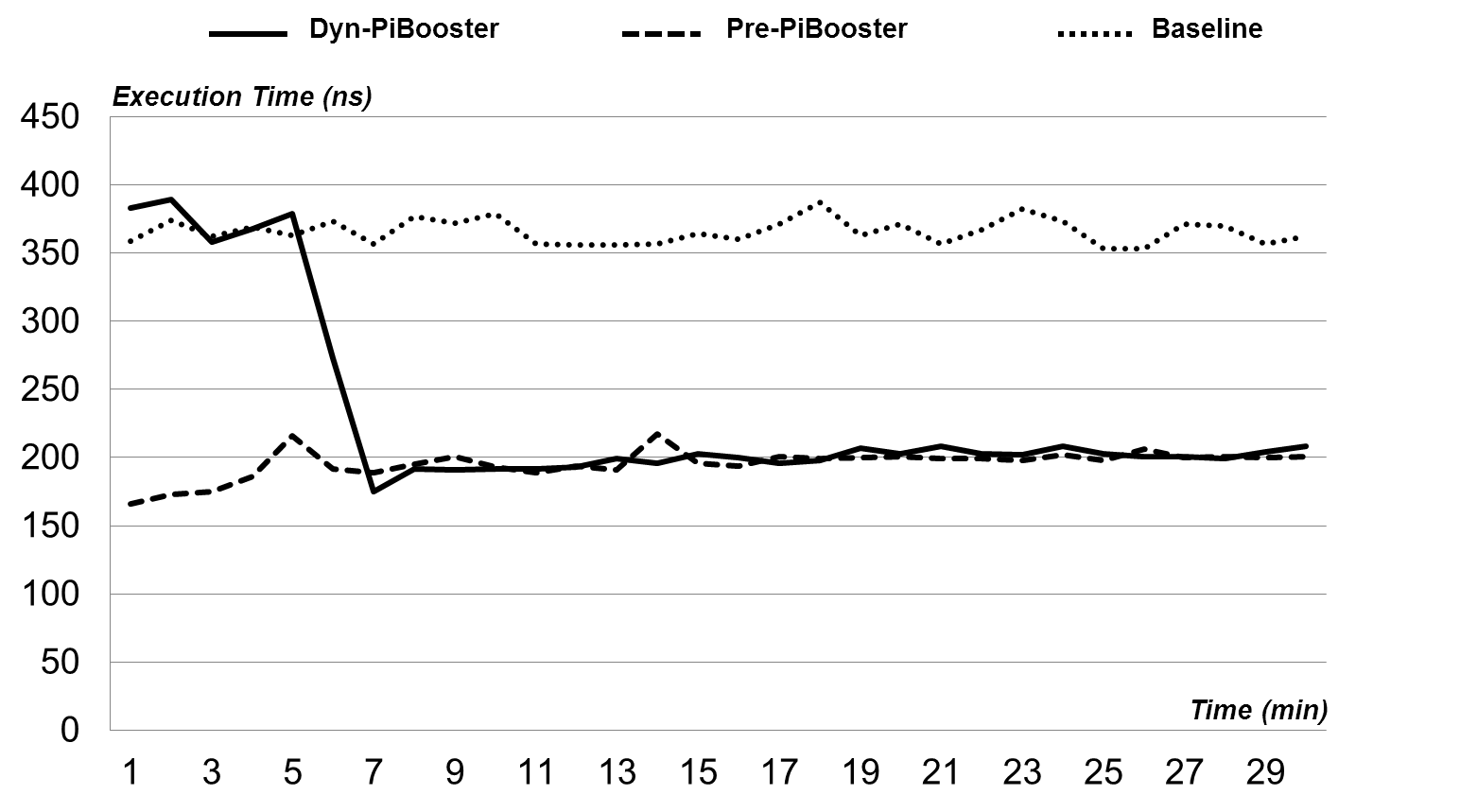}
        \caption{Execution time of L1 deallocation is reduced by 47\%.}
        \label{fig:l1b}
    \end{subfigure}
    \caption{The execution time of L1 (i.e., the top level) allocation and deallocation. The \name in the \dynname group can quickly enter the stable state in 2 minutes.}
    \label{fig:PGDtime}
\end{figure*}

\begin{figure*}[t!]
    \centering
    \begin{subfigure}[t]{0.5\textwidth}
        \centering
        \includegraphics[height=2.0in]{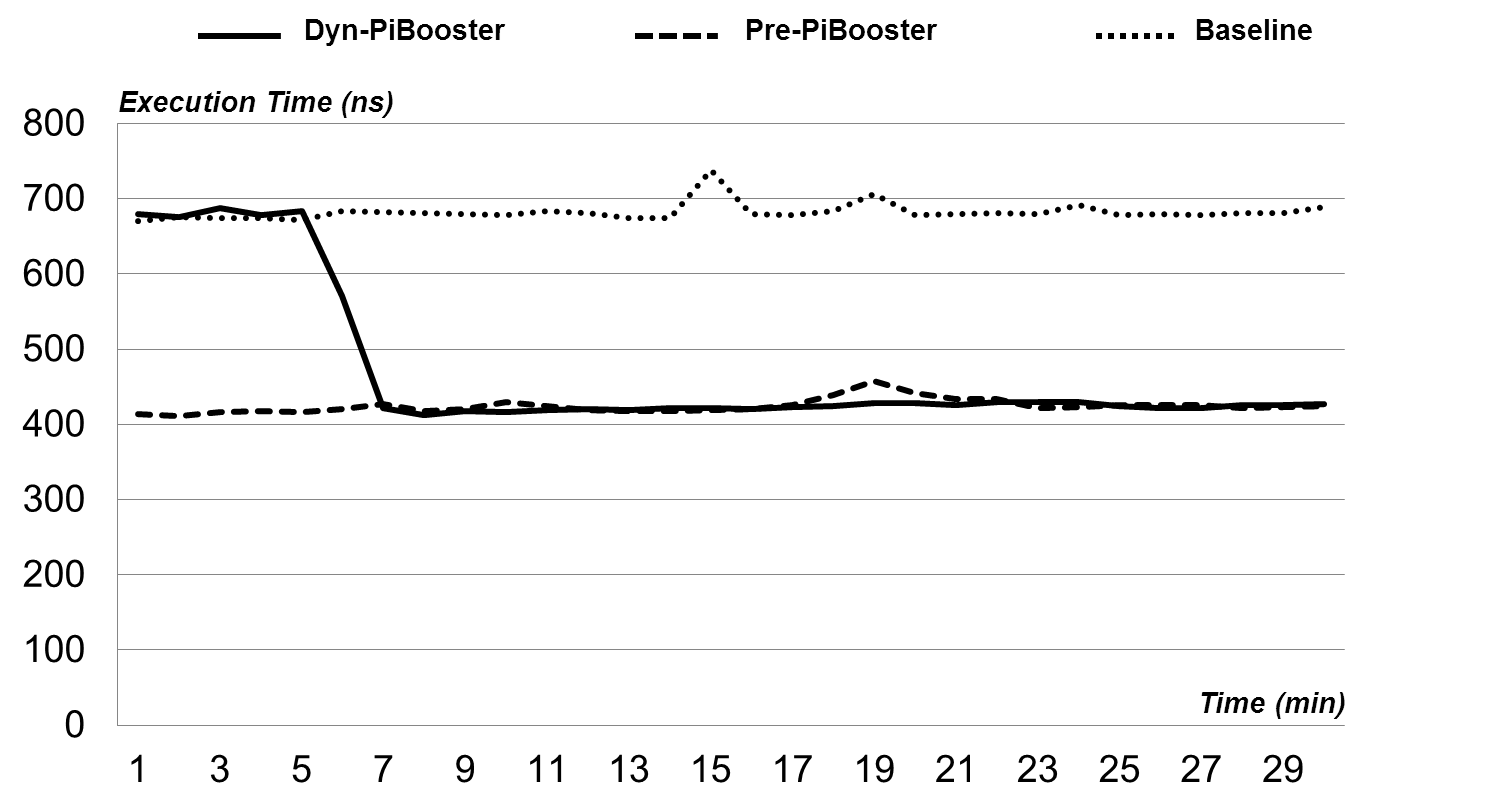}
        \caption{Execution time of L2 allocation is reduced by 38\%.}
        \label{fig:l2a}
    \end{subfigure}%
    ~
    \begin{subfigure}[t]{0.5\textwidth}
        \centering
        \includegraphics[height=2.0in]{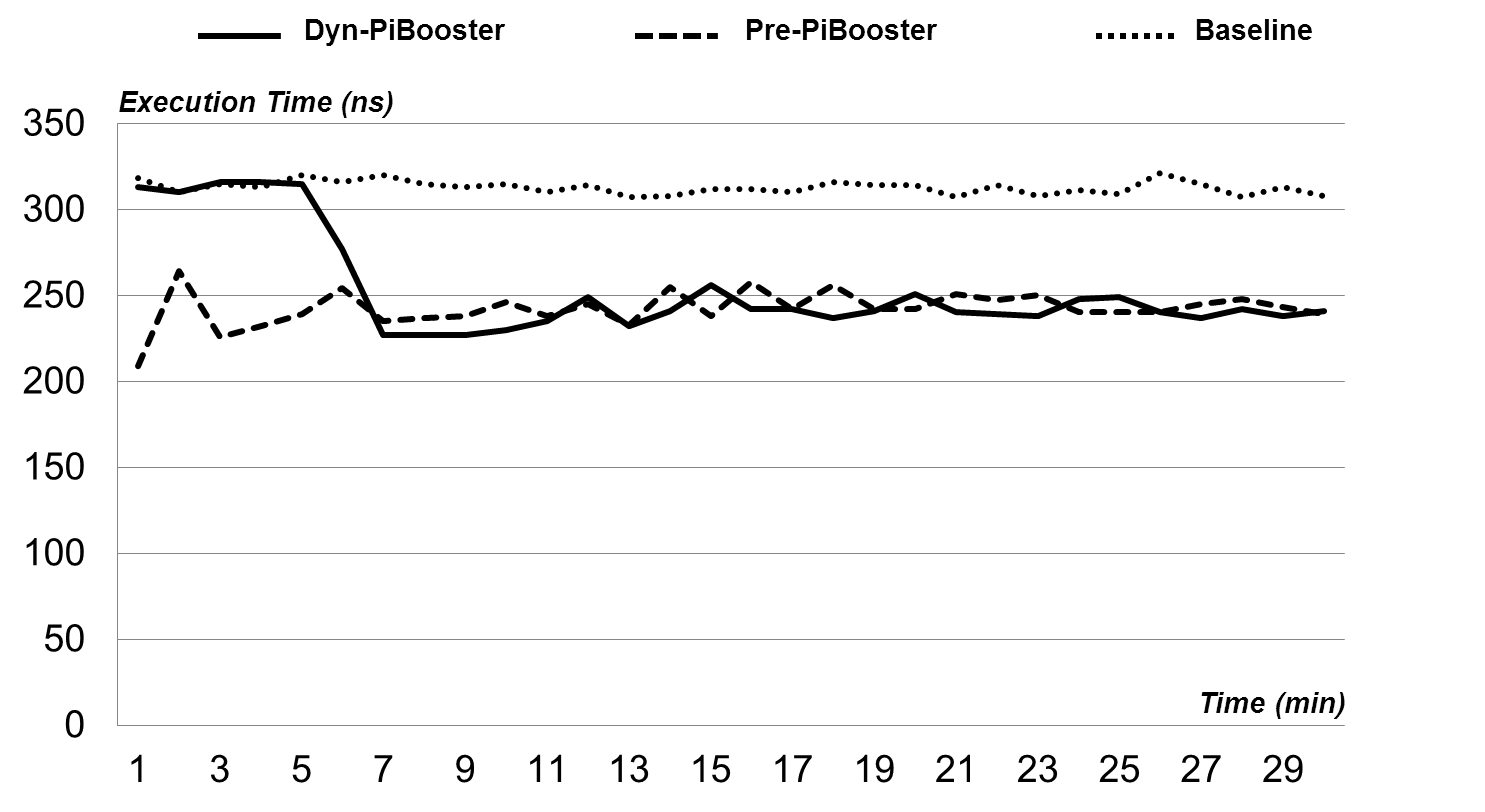}
        \caption{Execution time of L2 deallocation is reduced by 22\%.}
        \label{fig:l2b}
    \end{subfigure}
    \caption{The execution time of L2 allocation and deallocation. The \name in the \dynname group can quickly enter the stable state in 2 minutes.}
    \label{fig:PMDtime}
\end{figure*}

\begin{figure*}[t!]
    \centering
    \begin{subfigure}[t]{0.5\textwidth}
        \centering
        \includegraphics[height=2.0in]{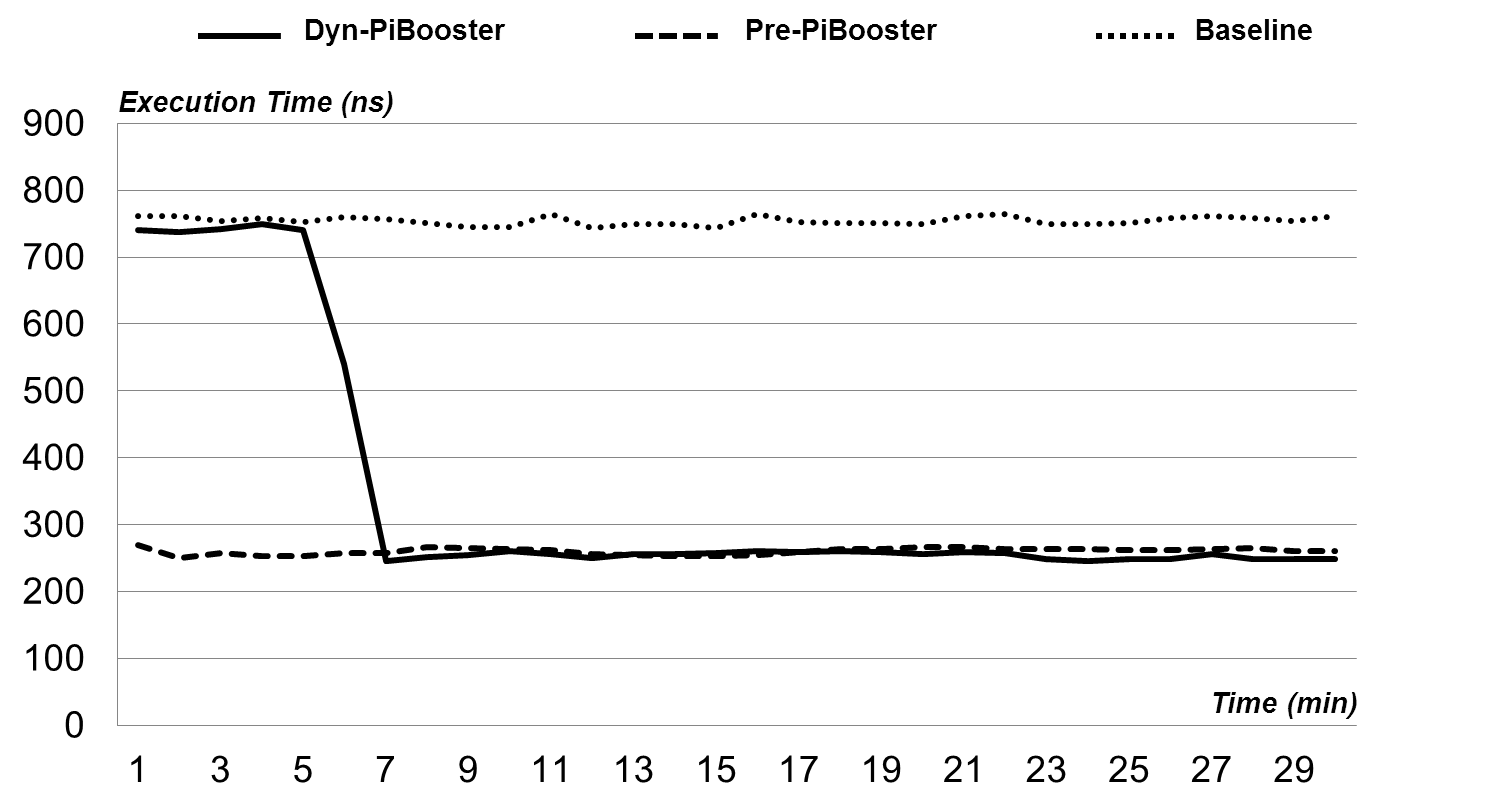}
        \caption{Execution time of L3 allocation is reduced by 65\%.}
        \label{fig:l3a}
    \end{subfigure}%
    ~
    \begin{subfigure}[t]{0.5\textwidth}
        \centering
        \includegraphics[height=2.0in]{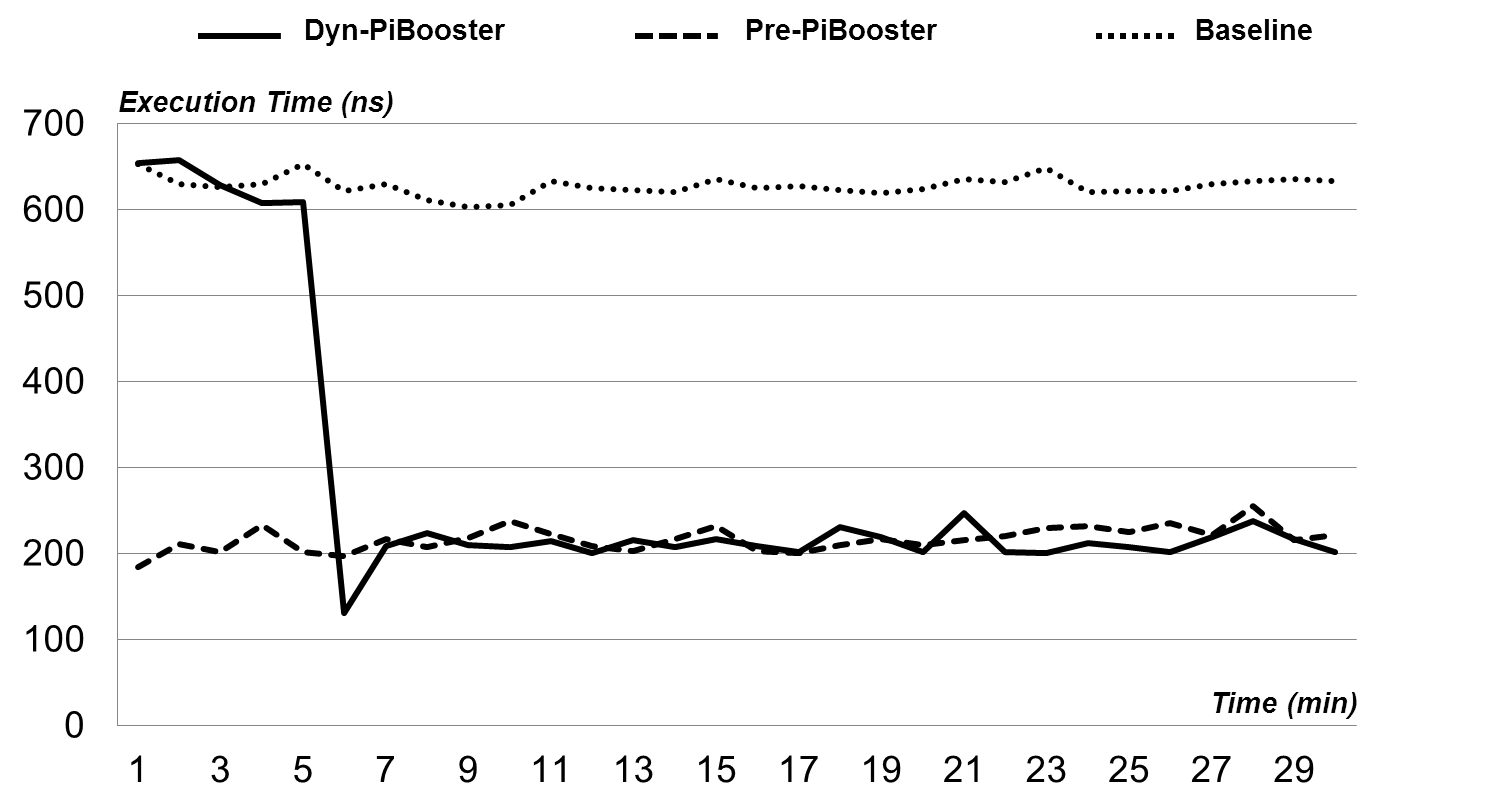}
        \caption{Execution time of L3 deallocation is reduced by 65\%.}
        \label{fig:l3b}
    \end{subfigure}
    \caption{The execution time of L3 (i.e., the bottom level) allocation and deallocation. The \name in the \dynname group can quickly enter the stable state in 2 minutes.}
    \label{fig:PTEtime}
\end{figure*}

\subsubsection{Execution Time Measurement}
As there are three-level of the guest page table, and each level has its (de)allocation functions, e.g., pgd\_alloc and pgd\_free for the L1 (top) level.
In order to clearly observe the changes of the CPU usage, we measure them separately.
We continuously measure them in 30 minutes, and calculate the average execution time of each function in each one minute.
Note that in the \dynname group, we enable the \name 5 minutes after the workload starts to run.
That is why the measurements of the \dynname group in the first 5 minutes are almost overlapped with the baseline results.

As shown in Figure~\ref{fig:PGDtime}, \ref{fig:PMDtime}, and \ref{fig:PTEtime}, the execution time on average in the \prename group, for a pair of allocation and deallocation of the three-level page table, from top to bottom are (622, 196), (424, 242) (260, 217) in nanoseconds, while in the baseline group, the corresponding execution time on average are (944, 366), (682, 313), (755, 627) in nanoseconds. Correspondingly, the improvements are (34\%, 47\%), (38\%, 22\%) and (65\%, 65\%), from top to bottom.
Putting them together, the page table allocations and deallocation can be improved by 45\% and 55\% on average respectively.
The results also indicate that the \name in the \dynname group and the \prename group can achieve the same performance improvements.
The only difference is that the \name in the \dynname group needs a transitional period of 2 minutes to be stable.

\mypara{The Worst Case}
When the \name starts to work, the guest kernel always first invokes the \cache for the page table allocations.
If the cached semi-writable pages cannot satisfy the requirements (a.k.a., cache miss), the \cache would have to allocate writable pages from the existing memory allocators following the traditional paths,.
In this case, the execution time is the traditional execution path plus the path in the \cache.
As a result, the execution time of the page table allocation is even longer than that of the baseline (Figure~\ref{fig:overview}a).
However, the overhead introduced by the \name path is negligible, as the control flow will immediately return when the cache is empty.
More specifically, the overhead consists of the function invocation, the stack adjustments and the checking logic of both the cache list and the page type.
Putting them together, the overhead are less than 20 instructions.
Fortunately, the worst case does not often occur. According to our observations in the \dynname group, the number of the worst cases is $348$, out of $198990$ allocation requests in 30 minutes.
Note that the page-table deallocation is always successful in a constant time.



\begin{table}[!ht]
\footnotesize
\begin{center}
\begin{tabular}{|l|l|l|}
\hline
{\textbf{Levels of Page Table}} & {\textbf{\prename (\#)}} & {\textbf{\dynname (\#)}} \\ \hline
L1 & $5$  & $4$ \\ \hline
L2 & $26$ & $20$ \\ \hline
L3 & $145$ & $136$ \\ \hline
Total & $176$ & $160$ \\ \hline
\end{tabular}
\end{center}
\caption{Cache usages in both groups of \prename and \dynname are small, occupying 176 pages and 160 pages, respectively.}
\label{tab:PGpool}
\end{table}

\subsubsection{Memory Usage Measurement}
To clearly observe the page usage in each level, we measure the numbers of the cached pages in three levels and the results are listed in Table \ref{tab:PGpool}.
In the \prename group the \cache has $5$, $26$ and $145$ semi-writable pages ready for allocations, totally occupying $704KB$, which is quite similar to the \dynname group, where the cache has $160$ in total, occupying $640KB$. As a result, the \cache in both groups consume an insignificant memory usage, always less $1MB$.


\subsection{Macro-Benchmarks}
The purpose of the macro-benchmarks is to evaluate the effects of \name on the overall system.
All measurements are divided into two groups: 1) the baseline group with the default settings, and 2) the \name group with the \name enabled.

\begin{figure}[htp]
\centering
\includegraphics[width=0.4\textwidth]{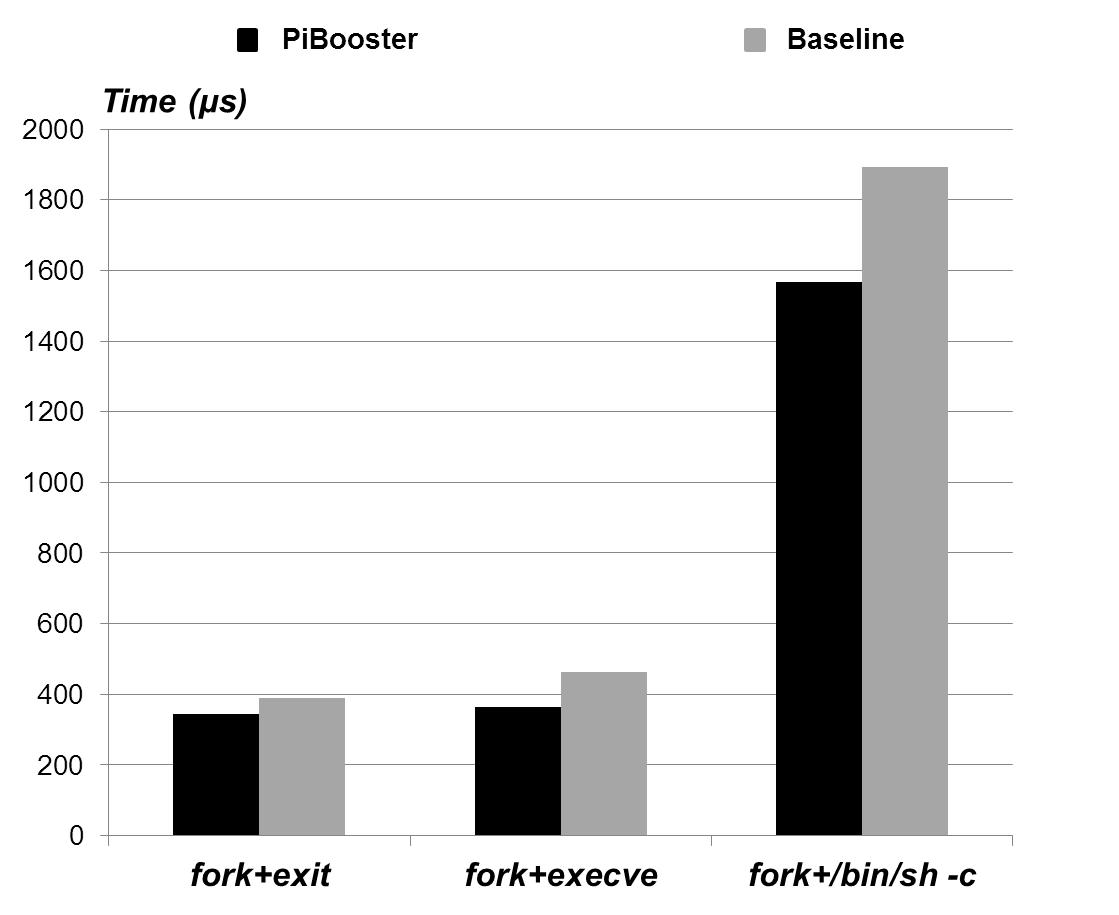} \\
\caption{The latencies on average is reduced by 11\%, 21\% and 17\%, from left to right.}
\label{fig:lmbench}
\end{figure}

\subsubsection{Latency of Process Creations and Exits}
Lmbench is a macro-benchmark tool for measuring the latencies of the process creations and exits (i.e., fork+exit, fork+execve, fork+/bin/sh -c), shown in Figure~\ref{fig:lmbench}.
The Lmbench is configured using the default parameters, except for the parameters of processor MHz and memory range.
In our experiment platform, the CPU frequency is $3.2$ GHz, memory range is set as $1024$ MB to save measurement time.
As illustrated in Figure~\ref{fig:lmbench}, the processes of \emph{fork+exit}, \emph{fork+execve} and \emph{fork+/bin/sh -c} in the \name group costs $344$, $365$, and $1567$ in microseconds, $11$\%, $21$\%, $17$\% faster than the ones in the baseline group.
We believe that the improvement will significantly benefit the workloads that rely on many temporary processes.

\begin{figure}[htp]
\centering
\includegraphics[width=0.5\textwidth]{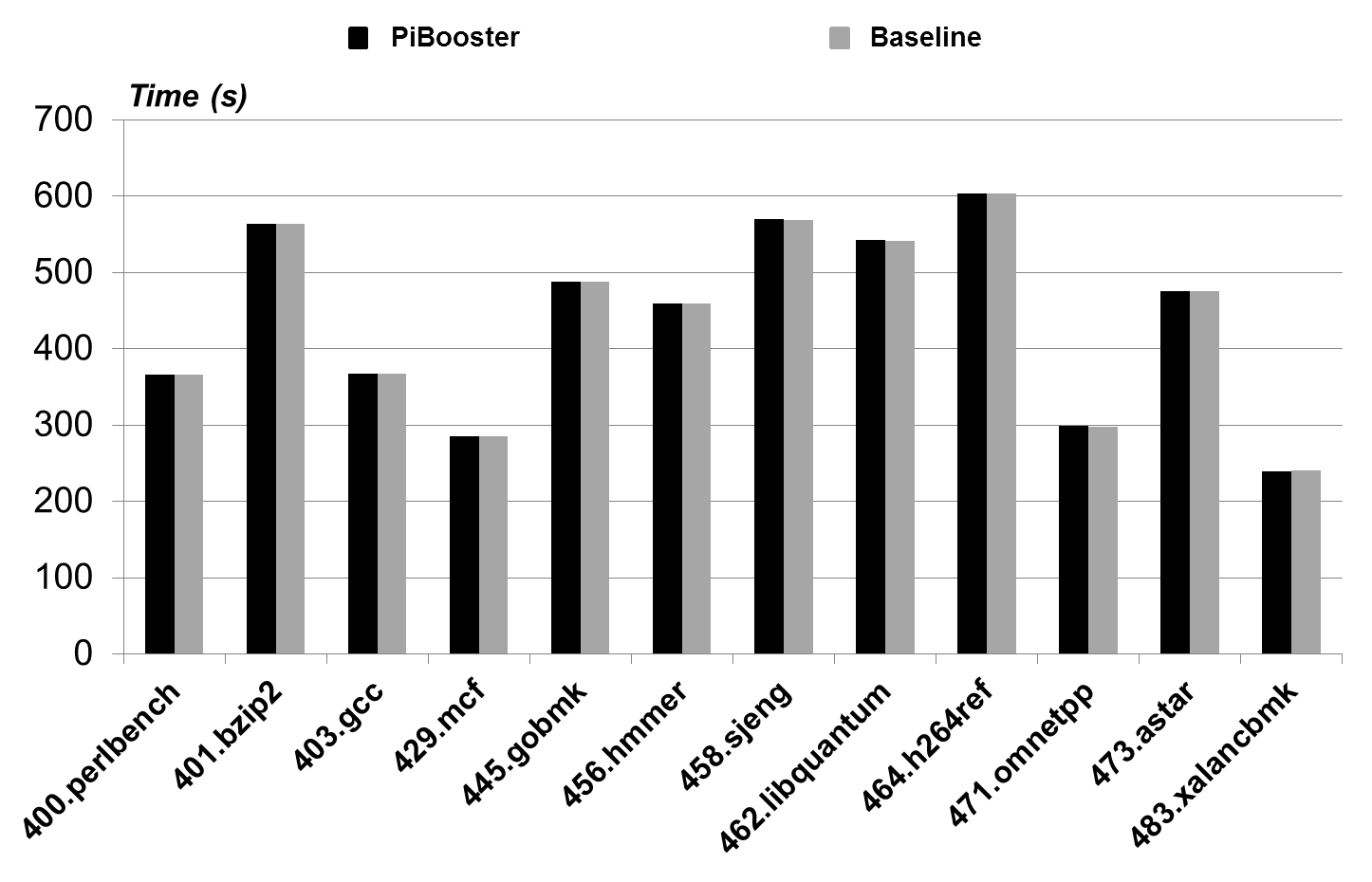} \\
\caption{The improvements among all the benchmarks are within 0.42\%, and the \name has no negative impact on the system performance.}
\label{fig:spec}
\end{figure}

\subsubsection{SPECINT}
SPECint2006~\cite{specint} is an industry standard benchmark intended for measuring the performance of the CPU and memory.
In our experiment, the tool version is SPECint 2006 v1.2, which has $12$ benchmarks in total and they are all invoked with the configuration file \emph{linux64-ia32-gcc43+.cfg}.
All measurement results are listed in Figure~\ref{fig:spec}.
Among the benchmarks, all benchmark tools in the \name group produce the same or a little better performance results, indicating that the \name has no negative effect on system performance.
The maximum improvement is about 0.42\%, which is produced by the \emph{483.xalancbmk}.

\subsubsection{I/O Performance Measurements}
As we know, IOTLB is used to accelerate the DMA address translation to achieve better performance for I/O devices.
Therefore, if there are many IOTLB misses caused by frequent IOTLB flushes, they will introduce negative effects on the I/O performance.
However,  rIOMMU~\cite{malka2015riommu} claims that the overhead caused by walking the IOMMU page tables due to IOTLB misses is so negligible that cannot be measured in the \emph{netperf}, because the main latency induced by I/O interrupt processing and the TCP/IP stack is several orders of magnitude larger than that of walking the page tables.
In addition, Nadav Amit et al.~\cite{amit2012iommu} also has similar statements.

In this paper, we test the network I/O and disk I/O performance under regular circumstances, and use these experiments to revisit of the problem between the IOTLB misses and I/O performance.
We measured the network I/O using \emph{netperf} tool. Specifically, we have two machines that are directly connected through the Ethernet cable.
The client on one machine sends a bulk of TCP packets to the server on another machine. Note that the emulated workload is also enabled on the client machine to trigger IOTLB flushes.
The sending buffer is $16KB$ and the test lasts 60 seconds.
The measurement results of the server are listed in Table~\ref{tab:netperf}.
By comparing both values of $\mu$ and $\sigma$, we find that there is no detectable improvement on the network I/O.

Similarly, we also test the disk I/O using \emph{lmbench}. The results indicate that the disk I/O speed remains the same.
The above two experiments as new evidences support the observations in ~\cite{amit2012iommu, malka2015riommu}.

\begin{table}[!ht]
\footnotesize
\begin{center}
\begin{tabular}{|l|l|l|}
\hline
{\textbf{Throughput (\emph{Mbps})}} & {\textbf{\name}} & {\textbf{Baseline}}    \\ \hline
Range & $87.880-88.010$ & $87.880-87.950$ \\ \hline
Arithmetic Mean ($\mu$)  &  $87.926$ & $87.913$ \\ \hline
Standard Deviation ($\sigma$) &  $0.028$ & $0.021$ \\ \hline
\end{tabular}
\end{center}
\caption{The netperf results of network I/O indicate that the overhead introduced by the IOTLB misses is negligible.}
\label{tab:netperf}
\end{table}

In fact, the effects of the IOTLB misses are measurable, it relies on an extremely high-speed I/O device, e.g., Intel's I/O Acceleration Technology~\cite{lauritzenintel}, or a newly designed IOMMU.
For instance, the rIOMMU project~\cite{malka2015riommu} establishes a high-performance setting with a newly designed IOMMU and the ibverbs library~\cite{ibverbsevaluation,kerr2011dissecting}.
We believe that the \name can effectively reduce the overhead in the above high-speed settings by eliminating the additional IOTLB misses, and we also plan to conduct experiments in the future.

\section{Discussion}\label{sec:dis}

\subsection{IOTLB Misses}
According to IOMMU specification~\cite{intelvt, amdvt}, a typical IOMMU is able to provide three types of IOTLB invalidation schemes, i.e., global invalidation, domain-selective invalidation, page-selective invalidation, which differ in granularity.
Specifically, the global invalidation will always invalidate all IOTLB entries as a whole.
The domain-selective invalidation only invalidates the selected VM domain's IOTLBs, whose performance is better than that of the global invalidation.
The page-selective invalidation that only invalidates the corresponding IOTLB entry could achieve the best performance, compared to the previous two schemes.

If the IOMMU is configured to do the global or domain-selective invalidation, an IOTLB flush will invalidate all IOTLB entries for at lease one domain, inevitably resulting in IOTLB misses for DMA address translations of the domain. However, if the IOMMU is working with the page-selective invalidation scheme, one IOTLB flush will only invalidate one IOTLB entry that may not be used immediately, which means that the current DMA request may not be affected, but the invalidation will badly affect DMA transfers in the near future .

\subsection{Threshold for \name Cache Shrinking}
Besides the command of explicitly shrinking the \cache by using the virtual files, the \cache could shrink itself according to the predefined threshold.
However, the threshold is likely to be different for different workload environments.
Thus, we should adjust the threshold according to the factors of the workload.
An interesting solution is to upgrade the \cache to allow itself to be aware of the workload updates and adjust the threshold accordingly.
It is not that easy to propose the self-adaption algorithm as there are many factors affecting the workload in a real system.
Another practical solution is to run a training tool in the guest kernel to calculate the number and the proportion of the needed semi-writable pages and the page table pages.
However, this approach requires that the workload is relatively stable, otherwise the end user has to do the training over and over again.
Fortunately, if the workload is updated in a regular way and the update pattern is known to the end user, the end user could use a shell script to automatically update the threshold with the help of the exported virtual files.


\subsection{\name Cache on the Bare-metal OS}
We believe that the design of the \cache could benefit the page table allocations and deallocations on the bare-metal OSes that work directly on hardware.
The usage pattern of the page table pages in paravirtual environment is similar if not the same to the one on the bare-metal OS.
Based on this feature, the deallocated page table pages are likely to be used in the near future by newly created processes.
By caching the deallocated page-table pages, the \cache could quickly response to the upcoming allocation requests, without the need to invoke the system allocators every time.
In the future, we plan to port the \cache onto a bare-metal OS, such as Linux, and fully evaluate its benefits.

\section{Related Work} \label{sec:related}
\mypara{Guest Page Table Protection}
In paravirtualization, the guest page table becomes the security critical data structure.
To protect the integrity of the page table and allow legitimate updates, Xen~\cite{barham2003xen} sets the guest page table read-only and validates every update of guest page tables so as to prevent any malicious access.
However, the protected guest page table is still writable for DMA requests~\cite{disaggregation,adams2006comparison}.
To fix this gap, the hypervisor has to enable the I/O virtualization (AMD-Vi~\cite{amdvt} or Intel VT-d~\cite{intelvt}) technology, preventing any DMA access to the guest page table.
In this paper, we do not break or downgrade the page-table based security, instead we keep its security and accelerate the performance in the page table management.

\mypara{IOTLB Misses Reduction}
There are some existing approaches~\cite{amit2012iommu, malka2015riommu, willmann2008protection} analyzing and reducing the negative effects due to the IOTLB misses.
Amit et al.~\cite{amit2012iommu} firstly analyze the role of the IOTLB in DMA operations and quantify the performance overhead of IOTLB misses. Then they present new strategies of both software and hardware enhancements to reduce IOTLB miss rate in order to facilitate DMA address translation. rIOMMU~\cite{malka2015riommu} re-designs the architecture of IOMMU to achieve high performance in DMA transactions, during which the IOTLB misses are also largely reduced. Willmann et al.~\cite{willmann2008protection} proposes new strategies for Xen to re-configure the addressing mode of IOMMU, resulting in fewer IOTLB misses.
Different from the previous approaches that attempt to reduce the overall IOTLB misses, our approach mainly focuses on eliminating the additional IOTLB misses introduced by the security validations (i.e., the DMA validation) during the guest page table allocations and deallocations.

\mypara{Other I/O Performance Improvements}
There are many schemes trying to improve the I/O performance in different directions.
The approaches~\cite{menon2006optimizing,4734994,santos2008bridging} aim to improve the network throughput by optimizing the paravirtual I/O model on network.
In CDNA~\cite{cdna}, the authors propose a method for concurrent and direct network access for virtual machines.
Ongaro et al.~\cite{ongaro2008scheduling} study the impacts of guest scheduling on guest I/O performance by concurrently running different combinations of processor-intensive, bandwidth-intensive and latency-sensitive workloads. The approaches in~\cite{gordon2012towards,har2013efficient} attempt to reduce the number of VM exits. The studies in~\cite{liao2008software,liu2009virtualization,shalev2010isostack,landau2011splitx,xu2013vturbo} accelerate I/O performance by  designating cores to specific uses.
Several studies, like ELI~\cite{eli} and vIC~\cite{vic} attempt to reduce the overhead of I/O interrupts in virtual environments.



\section{Conclusion} \label{sec:con}
The paravirtual guest OS has two important performance issues in page table management : 1) the long execution paths of page table (de)allocations and 2) the additional IOTLB flushes introduced by the DMA validations.
In this paper, we proposed the \name system to address the above problems.
We shortened the execution paths of the page table (de)allocations by introducing the \cache, which could quickly response to the page table (de)allocation requests.
We introduced a fine-grained validation scheme, which successfully eliminated all additional IOTLB flushes and saved the time cost of enforcing the DMA validations, which further reduced the execution paths.
We implemented a prototype of the \name and fully evaluated its performance in micro- and macro-benchmarks.
The micro experiment results indicated that \name could \emph{completely eliminate} the additional IOTLB flushes in the workload stable environments, and effectively reduced (de)allocation time of the page table by 47\% on average.
The macro benchmarks showed that the latencies of the process creations and exits were expectedly reduced by 16\% on average.
Moreover, the \emph{SPECINT}, \emph{lmbench} and \emph{netperf} results indicated that \name had \emph{no} negative impacts on CPU computation, network I/O, and disk I/O.

\ifCLASSOPTIONcaptionsoff
  \newpage
\fi

{\footnotesize \bibliographystyle{IEEEtranS}
\bibliography{main}}

\begin{thebibliography}{10}
\providecommand{\url}[1]{#1}
\csname url@samestyle\endcsname
\providecommand{\newblock}{\relax}
\providecommand{\bibinfo}[2]{#2}
\providecommand{\BIBentrySTDinterwordspacing}{\spaceskip=0pt\relax}
\providecommand{\BIBentryALTinterwordstretchfactor}{4}
\providecommand{\BIBentryALTinterwordspacing}{\spaceskip=\fontdimen2\font plus
\BIBentryALTinterwordstretchfactor\fontdimen3\font minus
  \fontdimen4\font\relax}
\providecommand{\BIBforeignlanguage}[2]{{%
\expandafter\ifx\csname l@#1\endcsname\relax
\typeout{** WARNING: IEEEtranS.bst: No hyphenation pattern has been}%
\typeout{** loaded for the language `#1'. Using the pattern for}%
\typeout{** the default language instead.}%
\else
\language=\csname l@#1\endcsname
\fi
#2}}
\providecommand{\BIBdecl}{\relax}
\BIBdecl

\bibitem{x86-pv-model}
http://wiki.xenproject.org/wiki/X86\_Paravirtualised\_Memory Management.

\bibitem{ibverbsevaluation}
http://www.scalalife.eu/book/export/html/434.

\bibitem{adams2006comparison}
K.~Adams and O.~Agesen, ``A comparison of software and hardware techniques for
  x86 virtualization,'' \emph{ACM Sigplan Notices}, vol.~41, no.~11, pp. 2--13,
  2006.

\bibitem{vic}
I.~Ahmad, A.~Gulati, and A.~Mashtizadeh, ``vic: Interrupt coalescing for
  virtual machine storage device io,'' in \emph{2011 USENIX Annual Technical
  Conference (USENIX ATC’11)}, 2011, p.~45.

\bibitem{amdvt}
AMD, ``Secure virtual machine architecture reference manual,'' Dec. 2005.

\bibitem{amit2012iommu}
N.~Amit, M.~Ben-Yehuda, and B.-A. Yassour, ``Iommu: Strategies for mitigating
  the iotlb bottleneck,'' in \emph{Computer Architecture}.\hskip 1em plus 0.5em
  minus 0.4em\relax Springer, 2012, pp. 256--274.

\bibitem{XEN-SOSP03}
P.~Barham, B.~Dragovic, K.~Fraser, S.~Hand, T.~Harris, A.~Ho, R.~Neugebauer,
  I.~Pratt, and A.~Warfield, ``Xen and the art of virtualization,'' in
  \emph{SOSP '03: Proceedings of the nineteenth ACM symposium on Operating
  systems principles}.\hskip 1em plus 0.5em minus 0.4em\relax New York, NY,
  USA: ACM, 2003, pp. 164--177.

\bibitem{barham2003xen}
------, ``Xen and the art of virtualization,'' \emph{ACM SIGOPS Operating
  Systems Review}, vol.~37, no.~5, pp. 164--177, 2003.

\bibitem{slaballocator}
J.~Bonwick \emph{et~al.}, ``The slab allocator: An object-caching kernel memory
  allocator.'' in \emph{USENIX summer}, vol.~16.\hskip 1em plus 0.5em minus
  0.4em\relax Boston, MA, USA, 1994.

\bibitem{specint}
S.~P.~E. Corporation, ``Spec cint2006,'' http://www.spec.org/.

\bibitem{eli}
\BIBentryALTinterwordspacing
A.~Gordon, N.~Amit, N.~Har'El, M.~Ben-Yehuda, A.~Landau, A.~Schuster, and
  D.~Tsafrir, ``Eli: Bare-metal performance for i/o virtualization,'' in
  \emph{Proceedings of the Seventeenth International Conference on
  Architectural Support for Programming Languages and Operating Systems}, ser.
  ASPLOS XVII.\hskip 1em plus 0.5em minus 0.4em\relax New York, NY, USA: ACM,
  2012, pp. 411--422. [Online]. Available:
  \url{http://doi.acm.org/10.1145/2150976.2151020}
\BIBentrySTDinterwordspacing

\bibitem{gordon2012towards}
A.~Gordon, N.~Har'El, A.~Landau, M.~Ben-Yehuda, and A.~Traeger, ``Towards
  exitless and efficient paravirtual i/o,'' in \emph{Proceedings of the 5th
  Annual International Systems and Storage Conference}.\hskip 1em plus 0.5em
  minus 0.4em\relax ACM, 2012, p.~10.

\bibitem{har2013efficient}
N.~Har'El, A.~Gordon, A.~Landau, M.~Ben-Yehuda, A.~Traeger, and R.~Ladelsky,
  ``Efficient and scalable paravirtual i/o system.'' in \emph{USENIX Annual
  Technical Conference}, 2013, pp. 231--242.

\bibitem{intelvt}
Intel, ``Intel virtualization technology for directed i/o,'' Sep. 2007.

\bibitem{kerr2011dissecting}
G.~Kerr, ``Dissecting a small infiniband application using the verbs api,''
  \emph{arXiv preprint arXiv:1105.1827}, 2011.

\bibitem{buddyallocator}
K.~C. Knowlton, ``A fast storage allocator.'' \emph{Communications of the ACM},
  no.~8, pp. 623--624, 1965.

\bibitem{landau2011splitx}
A.~Landau, M.~Ben-Yehuda, and A.~Gordon, ``Splitx: Split guest/hypervisor
  execution on multi-core.'' in \emph{WIOV}, 2011.

\bibitem{lauritzenintel}
K.~Lauritzen, T.~Sawicki, T.~Stachura, and C.~E. Wilson, ``Intel i/o
  acceleration technology improves network performance, reliability and
  efficiently. technology@ intel magazine, march 2005.''

\bibitem{liao2008software}
G.~Liao, D.~Guo, L.~Bhuyan, and S.~R. King, ``Software techniques to improve
  virtualized i/o performance on multi-core systems,'' in \emph{Proceedings of
  the 4th ACM/IEEE Symposium on Architectures for Networking and Communications
  Systems}.\hskip 1em plus 0.5em minus 0.4em\relax ACM, 2008, pp. 161--170.

\bibitem{liu2009virtualization}
J.~Liu and B.~Abali, ``Virtualization polling engine (vpe): using dedicated cpu
  cores to accelerate i/o virtualization,'' in \emph{Proceedings of the 23rd
  international conference on Supercomputing}.\hskip 1em plus 0.5em minus
  0.4em\relax ACM, 2009, pp. 225--234.

\bibitem{love2004linux}
R.~Love, S.~H.~W. Are, A.~C. Linus, L.~V. C.~U. Kernels, and B.~W. Begin,
  ``Linux kernel development second edition,'' 2004.

\bibitem{malka2015riommu}
M.~Malka, N.~Amit, M.~Ben-Yehuda, and D.~Tsafrir, ``riommu: Efficient iommu for
  i/o devices that employ ring buffers,'' in \emph{Proceedings of the Twentieth
  International Conference on Architectural Support for Programming Languages
  and Operating Systems}.\hskip 1em plus 0.5em minus 0.4em\relax ACM, 2015, pp.
  355--368.

\bibitem{menon2006optimizing}
A.~Menon, A.~L. Cox, and W.~Zwaenepoel, ``Optimizing network virtualization in
  xen,'' in \emph{Proc. USENIX Annual Technical Conference (USENIX 2006)},
  2006, pp. 15--28.

\bibitem{mochel2005sysfs}
P.~Mochel, ``The sysfs filesystem,'' in \emph{Linux Symposium}, 2005, p. 313.

\bibitem{disaggregation}
\BIBentryALTinterwordspacing
D.~G. Murray, G.~Milos, and S.~Hand, ``Improving xen security through
  disaggregation,'' in \emph{Proceedings of the Fourth ACM SIGPLAN/SIGOPS
  International Conference on Virtual Execution Environments}, ser. VEE
  '08.\hskip 1em plus 0.5em minus 0.4em\relax New York, NY, USA: ACM, 2008, pp.
  151--160. [Online]. Available:
  \url{http://doi.acm.org/10.1145/1346256.1346278}
\BIBentrySTDinterwordspacing

\bibitem{ongaro2008scheduling}
D.~Ongaro, A.~L. Cox, and S.~Rixner, ``Scheduling i/o in virtual machine
  monitors,'' in \emph{Proceedings of the fourth ACM SIGPLAN/SIGOPS
  international conference on Virtual execution environments}.\hskip 1em plus
  0.5em minus 0.4em\relax ACM, 2008, pp. 1--10.

\bibitem{4734994}
K.~K. Ram, J.~R. Santos, Y.~Turner, A.~L. Cox, and S.~Rixner, ``{Achieving 10
  Gb/s using safe and transparent network interface virtualization},'' in
  \emph{International Conference on Virtual Execution Environments}, 2009, pp.
  61--70.

\bibitem{santos2008bridging}
J.~R. Santos, Y.~Turner, G.~J. Janakiraman, and I.~Pratt, ``Bridging the gap
  between software and hardware techniques for i/o virtualization.'' in
  \emph{USENIX Annual Technical Conference}, 2008, pp. 29--42.

\bibitem{shalev2010isostack}
L.~Shalev, J.~Satran, E.~Borovik, and M.~Ben-Yehuda, ``Isostack-highly
  efficient network processing on dedicated cores.'' in \emph{USENIX Annual
  Technical Conference}, 2010, p.~5.

\bibitem{whitaker2002scale}
A.~Whitaker, M.~Shaw, and S.~D. Gribble, ``Scale and performance in the denali
  isolation kernel,'' \emph{ACM SIGOPS Operating Systems Review}, vol.~36,
  no.~SI, pp. 195--209, 2002.

\bibitem{willmann2008protection}
P.~Willmann, S.~Rixner, and A.~L. Cox, ``Protection strategies for direct
  access to virtualized i/o devices.'' in \emph{USENIX Annual Technical
  Conference}, 2008, pp. 15--28.

\bibitem{cdna}
P.~Willmann, J.~Shafer, D.~Carr, S.~Rixner, A.~L. Cox, and W.~Zwaenepoel,
  ``Concurrent direct network access for virtual machine monitors,'' in
  \emph{High Performance Computer Architecture, 2007. HPCA 2007. IEEE 13th
  International Symposium on}.\hskip 1em plus 0.5em minus 0.4em\relax IEEE,
  2007, pp. 306--317.

\bibitem{xu2013vturbo}
C.~Xu, S.~Gamage, H.~Lu, R.~R. Kompella, and D.~Xu, ``vturbo: Accelerating
  virtual machine i/o processing using designated turbo-sliced core.'' in
  \emph{USENIX Annual Technical Conference}, 2013, pp. 243--254.

\end{thebibliography}

\begin{IEEEbiography}[{\includegraphics[width=1in,height=1.25in,clip]{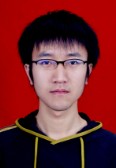}}]%
{Zhi Zhang} is a PhD student in the School of Computer Science and Engineering at the University of New South Wales. His research interests are in the areas of system security and virtualization. He received his bachelor degree from Sichuan University in 2011 and his master degree from Peking University in 2014.
\end{IEEEbiography}
\begin{IEEEbiography}[{\includegraphics[width=1in,height=1.25in,clip,keepaspectratio]{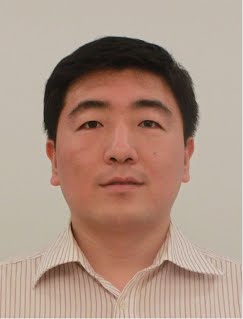}}]%
{Yueqiang Cheng} is a Staff Security Scientist at Baidu XLab America. He earned his PhD degree in School of Information Systems from Singapore Management University under the guidance of Professor Robert H. Deng and Associate Professor Xuhua Ding. 
His research interests are system security, trustworthy computing, software-only root of trust and software security.
\end{IEEEbiography}







\end{document}